# Modeling and control of an agile tail-sitter aircraft

Xinhua Wang †, Zengqiang Chen *, and Zhuzhi Yuan *

† Department of Mechanical and Aerospace Engineering,
Monash University, Melbourne, VIC, 3800, Australia Email: wangxinhua04@gmail.com

* Department of Automation, Nankai University, Tianjian, 300071, China

**Abstract:** This paper presents a model of an agile tail-sitter aircraft, which can operate as a helicopter as well as capable of transition to fixed-wing flight. Aerodynamics of the co-axial counter-rotating propellers with quad rotors are analysed under the condition that the co-axial is operated at equal rotor torque (power). A finite-time convergent observer based on Lyapunov function is presented to estimate the unknown nonlinear terms in co-axial counter-rotating propellers, the uncertainties and external disturbances during mode transition. Furthermore, a simple controller based on the finite-time convergent observer and quaternion method is designed to implement mode transition.

## 1 INTRODUCTION

This paper focuses on the design and control of an agile tail-sitter aircraft, where such an aircraft can not only taking off and landing vertically, but also flying forward with high speed in the same way as a conventional fixed wing aircraft.

Vertical take-off and landing (VTOL) aircrafts and fixed-wing airplanes have their advantages and shortcomings. Traditional aircrafts can take off and land vertically, but they cannot fly forward with high speed carrying large payloads [1-4]. On the other hand, conventional fixed-wing airplanes can fly forward with high speed and can carry large payloads. However, they cannot take off and land vertically, and appropriate runways are required.

There are some types of VTOL aircrafts with the ability of high-speed forward flight, such as manned aircrafts AV-8B Harrier [5] and F-35 [6]. These aircrafts are designed for specific environment mission. The reason these aircrafts can perform vertical take-off and landing is all due to their powerful engines with thrust vectoring or tilting jettubes. Such aircrafts use jet engines to provide the required thrust. Although they are powerful, the jet exhaust stream is very hot and harmful, and it can easily destroy the ground environment or inflict injuries to people nearby. These aircrafts are not suitable for use for many civil and rescue operations. Moreover, such VTOL aircrafts with jet engines are less efficient in hover than a conventional helicopter or a tilting-rotor aircraft of the same gross weight [7]. Importantly, the tilting mechanisms and control hardware increase the weight of the aircraft.

In recent years, there has been a considerable attention towards the propeller-pushing and flapping-wing aircrafts which can not only take off vertically, but also fly forward with high speed. A successful example includes V-22 aircraft [8] as well as tail-sitter designs [9-17]. The T-wing is a VTOL UAV that is capable of both wing-born horizontal flight and propeller born vertical mode flight including hover and descent. These aircrafts can be considered hybrid helicopter/fixed-wing aircrafts and have higher rotor disk loadings. In the tail-sitter aircrafts, a novel unmanned aircraft called SkyTote has been designed [18-22]. It was originally conceived as an airborne conveyor belt that would use a VTOL capability to minimize ground handling. The concept demonstrator is a 'tail-sitter' configuration and utilizes coaxial counter-rotating rotors. A relatively large cruciform tail provides directional control in the airplane modes as well as serving as a landing gear in the helicopter mode. However, a sufficiently large thrust force must be provided to complete mode transition. Such tail-sitter aircrafts are less efficient in hover than a conventional helicopter of the same gross weight but still are much more efficient than other VTOL aircrafts without rotating wings [7]. Furthermore, the attitude control is implemented based on the downwash flow generated by the coaxial counter-rotating propellers. Large size of the co-axial counter-rotating propellers is required. Alternatively, if the size of the co-axial counter-rotating propellers is restricted, the upward flying velocity of the aircraft should attain a given value which can provide the sufficient moments in level and vertical flying modes. This adds constraints to the types of the flying trajectories possible.

A tilt-fuselage aircraft was presented in [23] to keep the flying height invariant during mode transitions. It is a rotor-fixed wing aircraft with two free wings. During mode transition, the fuselage is tilted and free wings are kept at a given small angle of attack. However, it is difficult to analyze the aerodynamics of the tilting fuselage during mode transition, and moreover, the tilting structure is difficult to control.

In this paper, a novel agile tail-sitter aircraft is presented. Its tilt structure is based on a quad rotor. When the aircraft hovers, takes off or lands, control method of a quadrotor aircraft can be used directly [24, 25, 26]. During mode transition from hover to forward flight and vice versa, the tilt moments are generated by the force differential of the two pairs of rotors. The co-axial counter-rotating propellers provide the thrust. Comparing with the conventional tail-sitter aircraft, more agile maneuverability can be obtained. Aerodynamics of the counter-rotating propellers with quad rotors is analyzed under the condition that the co-axial is operated at equal rotor torque (power). In order to reconstruct the nonlinear terms in the relationship between the thrust and the rotational speed, the uncertainties and the external disturbances, a finite-time convergent observer is presented to estimate the unknown terms. Furthermore, the quad rotors increase the force efficiency of the co-axial counter-rotating propellers with respect to the two independent ones. A simple controller based on the observer and quaternion method is designed to implement mode transition. The flying modes transition is shown in Figure 1.

The outline of the paper is as follows. In section 2, we present the design of aircraft including the mechanical structure of aircraft. In Section 3, the mathematical model of aircraft is derived, working from first principles and basic aerodynamics. In section 4, observer design is proposed. In section 5, controller design is proposed. In section 6, desired trajectory during mode transition is described. Computational analysis and simulation experiments are presented in Section 7. The conclusions are provided in Section 8. The Appendix for the proofs of some theorems is in Section 9. In Section 10, list of symbols is shown.

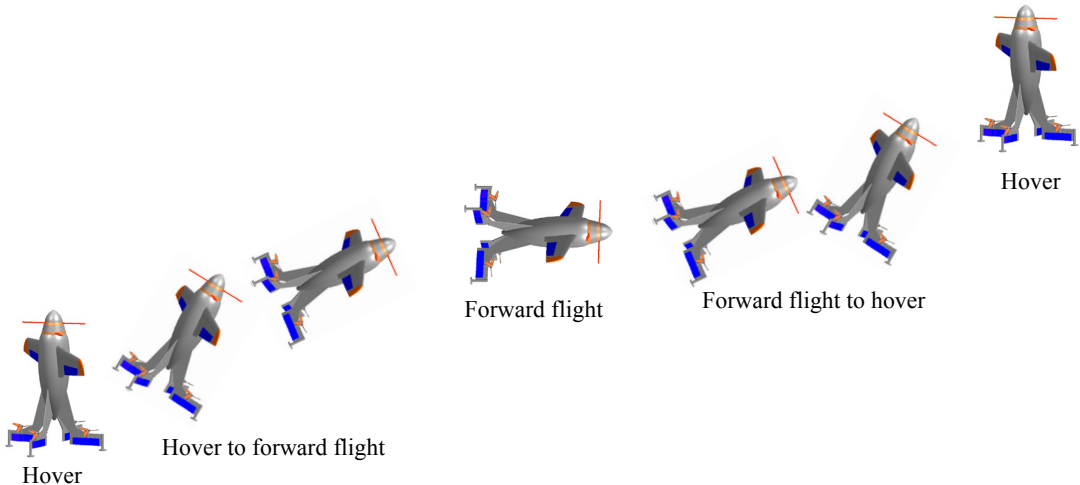

Figure 1 Transition to forward flight from hover and vice versa

## 2 AIRCRAFT DESIGN
### 2.1 Mechanical structure of the aircraft

A tail-sitter aircraft is presented in Figures 2 (a)-(d).

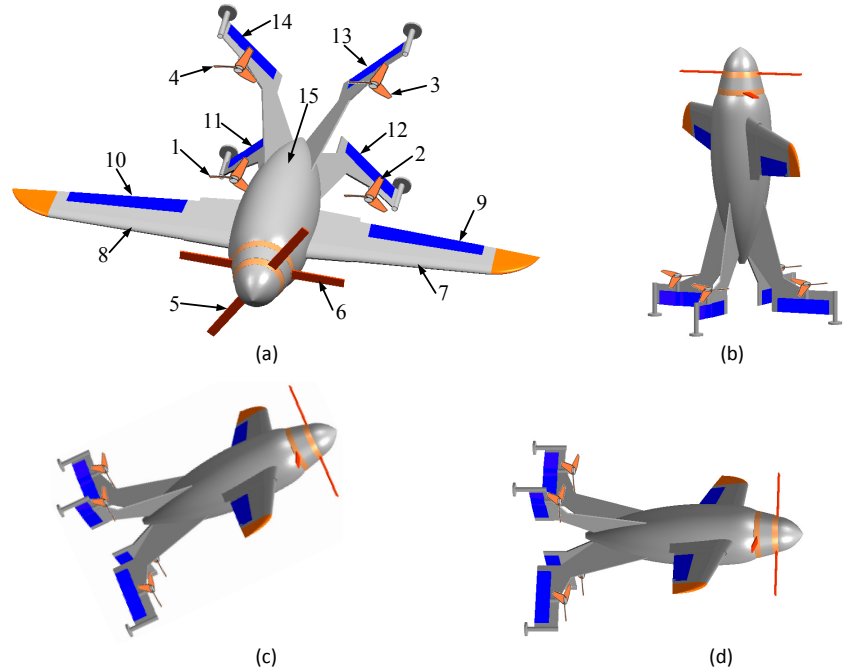

Figure 2 VTOL tail-sitter aircraft: (a) Structure of tail-sitter aircraft, (b) Hover, taking-off or landing mode, (c) Mode transition from hover to forward flight and vice versa, (d) Forward flight

**Notations:**
1: rotor 1, 2: rotor 2, 3: rotor 3, 4: rotor 4, 5: up propeller, 6: low propeller, 7: left fixed wing, 8: right fixed wing, 9: left aileron, 10: right aileron, 11: vane 1, 12: vane 2, 13: vane 3, 14: vane 4, 15: fuselage

*A. Flying modes*

When the aircraft is in VTOL flight or in hover (see Figure 2(b)), the thrusts generated by the propellers 5 and 6 with quad rotors 1-4 provide the required lift force. A control method for quad rotor aircraft can be used. The only difference is that the main lift force can be provided by the co-axial propellers 5 and 6, and the attitude regulation is provided by quad rotors 1-4.

For this aircraft, the yaw dynamics in forward flight correspond to the roll dynamics in hover, the pitch dynamics in forward flight correspond to the same dynamics in hover, and the roll dynamics in forward flight correspond to the yaw dynamics in hover. In the following, we select the dynamic angle names in forward flight for all the flying modes.

For transition from hover to horizontal flight, assuming that the aircraft is hovering (see Figure 2(b)), the aircraft is initially lifted by co-axial counter-rotating propellers with quad rotors 1-4. The thrusts generated by rotors 3 and 4 increase, at the same time the thrusts generated by rotors 1 and 2 decrease. Thus, the fuselage is tilted towards the horizontal, which in turn causes the horizontal speed of the aircraft to increase (see Figure 2(c)). With the regulation of the co-axial counter-rotating propellers 5 and 6, the fixed wings 7 and 8 obtain a given angle of attack in accordance with the relative wind. The gravity of the aircraft is counteracted mainly by the vertical force of the thrusts generated by co-axial counter-rotating propellers and quad rotors 1-4. The flying process is shown in Figure 1. Quad rotors 1-4 are controlled, the pitch angle changes from 90 degree to zero degree. With increasing horizontal speed, wings 7 and 8 develop lift. The aircraft soon transitions into horizontal flight in a fixed wing straight and level flight mode (see Figure 2(d)). During mode transition from hover to forward flight, roll, yaw and pitch dynamics are all controlled by quad rotors 1-4. The attitude control is similar to that of usual quad-rotor aircrafts. The only difference is that a torque amplifier for the reactive torque is designed for magnifying the roll moment, because the coefficient of the reactive torques generated by quad rotors 1-4 is very small.

For transition from horizontal flight to hover, the aircraft is controlled to climb up. The aircraft flies towards vertical (see Figure 1). This causes the horizontal speed of the aircraft to decrease and the vertical thrust vector gradually increases to overcome the gravity. Thus, the aircraft slows and performs transitions to hover.

*B. Analysis without the quad rotors 1-4*

The aircraft without quad rotors 1-4 cannot provide the sufficient pitch, roll and yaw torques (see Figure 3). The attitude control is implemented based on the downwash flow generated by the co-axial counter-rotating propellers. Therefore, a large size of the co-axial counter-rotating propellers is required. Furthermore, a sufficiently large thrust generated by the co-axial counter-rotating propellers is needed. Alternatively, if the size of the co-axial counter-rotating propellers is restricted, the upward flying velocity of the aircraft should attain a given value which can provide the sufficient torques by vanes 11-14. This restrains the types of the flying trajectories.

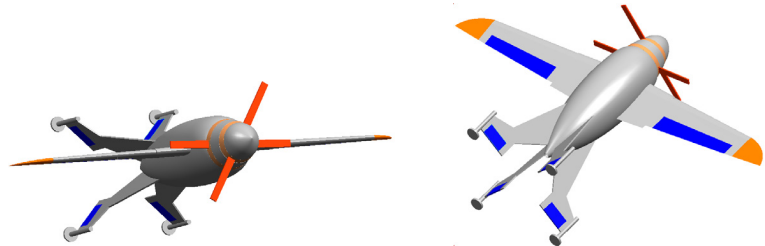

Figure 3 VTOL Tail-sitter aircraft without quad rotors 1-4

# 3 MATHEMATICAL MODEL

## 3.1 Mathematical model in hover

The forces and moments for the tail-sitter aircraft during hover, vertically takeoff and landing are shown in Figures 4 and 5.

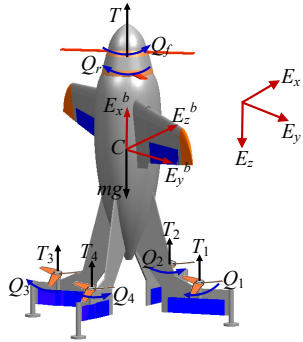 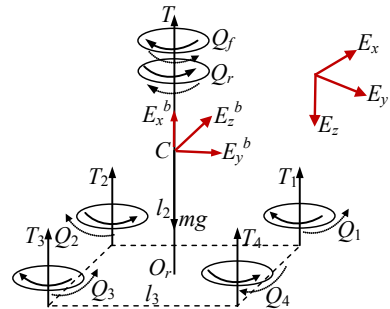

Figure 4 VTOL aircraft model in hover     Figure 5 Forces and torques of the aircraft in hover

The modeling and control is similar to normal quadrotor aircrafts during hovering, takeoff and landing. Here we will not discuss this case.

## 3.2 Mathematical model during mode transition

The forces and moments for the aircraft during mode transition are shown in Figures 6 and 7.

As the blades of quad rotors 1-4 rotate, they are subjected to drag forces which produce torques around the aerodynamic center. These moments act in opposite direction relative to the rotation rate of the rotor. The reactive torque generated, in free air, by the rotor due to rotor drag is small. It is difficult to regulate the roll dynamics during mode transition. Therefore, a torque amplifier for roll dynamics during mode transition is designed as shown in Figure 8. In Figure 8, when the rotors 1-4 rotate, the biases of the vanes 1-4 generate the forces $f_1, f_2, f_3$ and $f_4$. These forces can help increase the roll torque.

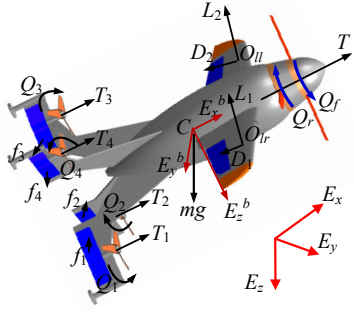
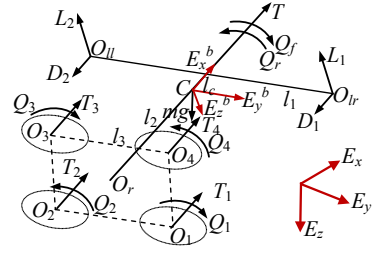

Figure 6 VTOL tail-sitter aircraft model during mode transition    Figure 7 Forces and torques of the aircraft during mode transition

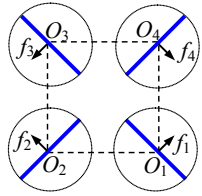

Figure 8 Torque amplifier for roll dynamics during mode transition

### 3.2.1 Coordinates and frames

In Figures 6 and 7, $C$ is the centre of gravity of the aircraft. Let $\Gamma = (E_x, E_y, E_z)$ denote the right handed inertial frame, and $\Lambda = (E_x^b, E_y^b, E_z^b)$ denote the frame attached to the aircraft's body whose origin is located at its center of gravity [27]. $P_\Lambda = (x_b, y_b, z_b)$ is the position of center of gravity relative to frame $\Lambda$. $\Theta_\Gamma = (\phi, \theta, \psi)^T \in \mathfrak{R}^3$ describes the aircraft orientation expressed in the classical yaw, pitch and roll angles (Euler angles), and $\Omega_\Gamma = (\dot\phi, \dot\theta, \dot\psi)^T$. We use $c_\theta$ for $\cos\theta$ and $s_\theta$ for $\sin\theta$. $R$ is the transformation matrix representing the orientation of the rotorcraft,

$$R = \begin{bmatrix} c_\theta c_\psi & c_\psi s_\theta s_\phi - s_\psi c_\phi & c_\psi s_\theta c_\phi + s_\psi s_\phi \\ c_\theta s_\psi & s_\psi s_\theta s_\phi + c_\psi c_\phi & s_\psi s_\theta c_\phi - c_\psi s_\phi \\ -s_\theta & s_\phi c_\theta & c_\phi c_\theta \end{bmatrix} \quad (1)$$

and the following relation holds: $V_\Gamma = R V_\Lambda$, where $V_\Gamma$ denotes the vector in frame $\Gamma$, and $V_\Lambda$ is its projection in frame $\Lambda$. Let $\alpha$ be the angle of attack of the fixed wing, and

$$\alpha = \theta - \arctan^{-1}(\dot z_b / \dot x_b) \quad (3)$$

Let $\beta$ be the sideslip angle, and

$$\beta = \arcsin^{-1}(\dot y_b / V_b) \quad (4)$$

where $V_b = \sqrt{\dot x_b^2 + \dot y_b^2 + \dot z_b^2}$.

### 3.2.2 Dynamical model

By defining $P_\Gamma = (x, y, z)$ and $\upsilon_\Gamma = (\dot x, \dot y, \dot z)$ as the position of center of gravity and the velocity relative to frame $\Gamma$, the equations of motion for a rigid body subjected to body force $F \in \mathfrak{R}^3$ and torque $\tau \in \mathfrak{R}^3$ applied at the center of mass and specified with respect to frame $\Gamma$, are given as follows:

$$\begin{aligned} \dot P_\Gamma &= \upsilon_\Gamma \\ m\dot\upsilon_\Gamma &= F - mgE_z \end{aligned} \quad (2)$$

$$J\dot{\Omega}_\Lambda + \Omega_\Lambda \times J\Omega_\Lambda = \tau \tag{3}$$

where $\Omega_\Lambda = [p_\Lambda \quad q_\Lambda \quad r_\Lambda]^T \in \Re^3$ denotes the angular velocity of the airframe expressed in frame $\Lambda$, $m \in \Re$ specifies the mass, and $J = \text{diag}\{J_{xb}, J_{yb}, J_{zb}\} \in \Re^{3\times 3}$ is an inertial matrix; $\dot{R} = RS(\Omega_\Lambda)$; The skew-symmetric matrix $S(\Omega_\Lambda)$ is defined as follow:

$$S(\Omega_\Lambda) = \begin{bmatrix} 0 & -r_\Lambda & q_\Lambda \\ r_\Lambda & 0 & -p_\Lambda \\ -q_\Lambda & p_\Lambda & 0 \end{bmatrix} \tag{4}$$

The relation between the angular velocity of the aircraft and the time derivative of the attitude angles is given by the following transformation

$$\Omega_\Lambda = [p_\Lambda \quad q_\Lambda \quad r_\Lambda]^T = \mathbb{Z}\Omega_\Gamma \tag{5}$$

where $\mathbb{Z}$ is the velocity transformation matrix and defined as

$$\mathbb{Z} = \begin{bmatrix} 1 & 0 & -s_\theta \\ 0 & c_\phi & s_\phi c_\theta \\ 0 & -s_\phi & c_\phi c_\theta \end{bmatrix} \tag{6}$$

Therefore, we obtain the following relation between the angular rate and the derivatives of the Euler angles:

$$\Omega_\Gamma = \begin{bmatrix} \dot{\phi} \\ \dot{\theta} \\ \dot{\psi} \end{bmatrix} = \mathbb{Z}^{-1}\Omega_\Lambda = \mathbb{Z}^{-1} \begin{bmatrix} p_\Lambda \\ q_\Lambda \\ r_\Lambda \end{bmatrix} \tag{7}$$

where

$$\mathbb{Z}^{-1} = \begin{bmatrix} 1 & \sin\phi\tan\theta & \cos\phi\tan\theta \\ 0 & \cos\phi & -\sin\phi \\ 0 & \sin\phi\sec\theta & \cos\phi\sec\theta \end{bmatrix} \tag{8}$$

The total external force $F$ consists of the thrust $F_c$ generated by the co-axial counter rotating propellers, the thrusts $F_r$ of the quad rotors, aerodynamic forces on the fixed wings $F_w$, aerodynamic forces on the fuselage $F_f$, and forces due to uncertainties and external disturbances $F_d$. These forces are expressed in body frame $\Lambda$, and they are transformed by $R$ to be expressed in the inertial frame $\Gamma$ as follows:

$$F = R(F_c + F_r + F_w + F_f + F_d) \tag{9}$$

The total moment $\tau$ consists of the moments created by the rotors $\tau_r$, moments created by the aerodynamic forces produced by the wings $\tau_w$, moments created by the gyroscopic effects of the propellers $\tau_{gyro}$ and moments due to the uncertainties and external disturbances $\tau_d$:

$$\tau = \tau_r + \tau_w + \tau_{gyro} + \tau_d \tag{10}$$

One of the drawbacks related to the use of the Euler angle system is the inherent singularity. This drawback can be avoided by using the quaternion representation [28-32], which is based upon the fact that any rotation of a rigid body can by described by a single rotation about a fixed axis [33]. This globally nonsingular representation of the orientation is given by the vector $(q, q_0)^T$ with

$$q = \hat{k}\sin(\gamma/2), q_0 = \cos(\gamma/2) \tag{11}$$

where $\gamma$ is the equivalent rotation angle about the axis described by the unit vector $\hat{k} = (\hat{k}_1, \hat{k}_2, \hat{k}_3)$, with the constant as follow:

$$q^T q + q_0^2 = 1 \tag{12}$$

Although the quaternion representation is nonsingular, it contains a sign ambiguity (i.e., $(q, q_0)$ and $(-q, -q_0)$ lead to the same orientation) which can be resolved by choosing the following differential equations [32]:

$$\dot{q} = \frac{1}{2}(S(q) + q_0 I)\Omega_\Lambda, \quad \dot{q}_0 = -\frac{1}{2}q^T \Omega_\Lambda \tag{13}$$

where $I$ is a 3×3 identity matrix, and $S(\cdot)$ has been defined in Eq. (4). Moreover, the relationship between the quaternion and the Euler angles can be written as

$$\begin{aligned}
q_0 &= c_{\phi/2}c_{\theta/2}c_{\psi/2} + s_{\phi/2}s_{\theta/2}s_{\psi/2} \\
q_1 &= s_{\phi/2}c_{\theta/2}c_{\psi/2} + c_{\phi/2}s_{\theta/2}s_{\psi/2} \\
q_2 &= c_{\phi/2}s_{\theta/2}c_{\psi/2} + s_{\phi/2}c_{\theta/2}s_{\psi/2} \\
q_3 &= c_{\phi/2}c_{\theta/2}s_{\psi/2} + s_{\phi/2}s_{\theta/2}c_{\psi/2}
\end{aligned} \tag{14}$$

and

$$\begin{aligned}
\phi &= \tan^{-1}\left(\frac{2(q_0 q_1 + q_2 q_3)}{1 - 2 q_1 q_1 + q_2 q_2}\right) \\
\phi &= \sin^{-1}(2(q_0 q_2 - q_3 q_1)) \\
\psi &= \tan^{-1}\left(\frac{2(q_0 q_3 + q_1 q_2)}{1 - 2(q_2 q_2 + q_3 q_3)}\right)
\end{aligned} \tag{15}$$

### 3.3 Fluent for the aerodynamic parameters

In the simulation, in order to obtain the parameters instead of the actual parameters in the wind tunnel test, we use Fluent software to simulate the flying environment. Fluent is one of the applications of computing fluid dynamics [34]. It uses finite-element method to calculate the motion of fluid field, and three steps are arranged to get the aerodynamic parameters [35].

*Step 1: Meshing the fluid field.* The parameters of fixed wings 7, 8 and fuselage 15 are obtained by using the 3-D simulation shown in Figure 9(a), the parameters of one of the co-axial counter-rotating propellers use the simulation shown in Figure 9 (b), and the parameters of the rotors 1-4 with fairings 11-14 use the simulation shown in Figure 9 (c).

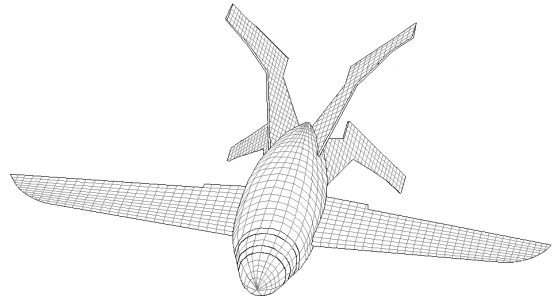

(a) Mesh of tail-sitter aircraft

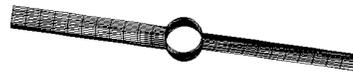 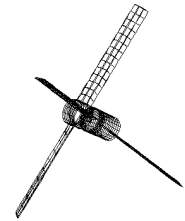

(b) Mesh of one of the co-axial counter-rotating propellers     (c) Mesh of the rotor with fairing

Figure 9 Model of fluent simulation

*Step 2: Fluent simulation*. The following conditions are set: the boundary condition, continuous equation, motion equation, energy equation, initial condition of the fluid field; the following constraints are selected: the type of the fluid field, viscid or invisicid, laminar or turbulence flow, k-epsilon, the algorithm to simulate the motion of fluid field. The parameter simulation results are shown in Figure 10.

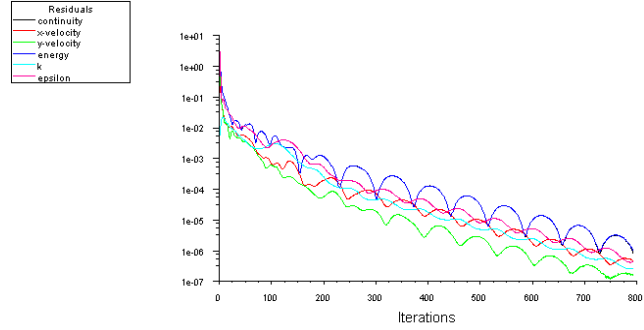

(a) Remains of the simulation target for the fixed wing and fuselage

(3-D velocity, energy, *k-s* model parameters)

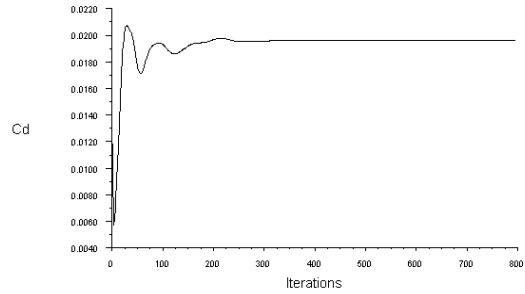

(b) Drag coefficient of fixed wing

Figure 10 Fluent simulation

If the curves in Figure 10(a) cannot converge to a preseted value, such as $10^{-6}$, then we should turn back to step 1, to make the mesh denser, and to repeat the step 2 until the curves converge to the preseted value.

*Step 3: Postprocessing*. After the Fluent simulation, we use the post processing tools in Fluent to get the needed parameters.

*1) The parameters of fixed wing*

The lift force and drag force generated by the fixed wings 7 and 8 are, respectively

$$\begin{aligned}
& L_i = 0.5 C_{Li} S \rho (\dot{x}_b^2 + \dot{z}_b^2), C_{Li} = C_{L0} + C_{L\alpha}\alpha + C_{L\delta_i}\delta_i \\
& D_i = 0.5 C_{Di} S \rho (\dot{x}_b^2 + \dot{z}_b^2), C_{Di} = C_{D_0} + C_{L_i}^2 / (\pi A_w e_w), e_w = 1.78(1 - 0.045 A_w^{0.68}) - 0.46 \\
& C_{L0} = 0.3137, C_{L\alpha} = 0.7025, C_{D0} = 0.00182, C_{L\delta i} = 0.1634, A_w = 6
\end{aligned} \quad (16)$$

where $i = 1, 2$; $S$ is the area of the half wing, $C_{L0}$ is the lift coefficient when the angle of attack $\alpha$ is equal to zero, $C_{L\alpha}$ is the lift coefficient due to the angle of attack $\alpha$, $\delta_i$ is the normal flap bias angle, and $C_{L\delta}$ is the lift coefficient due to the flap bias angle $\delta_i$. $A_w$ is aspect ratio of fixed wing. $e_w$ is the value of the Oswald's efficiency factor. The expression of lift and drag coefficients is considered as valid for low angle of attack, and the angle of attack could be higher during the initial mode transition at low speed.

A projection of lift and drag in the body frame is generated by the sideslip angle $\beta$. Here sideslip angle $\beta$ is assumed to be low enough to neglect this lateral effect. Alternatively, we can incorporate this effect into external uncertain force $F_d$. Then the aerodynamic forces on the fixed wings $F_w$ in body frame can be written as

$$F_w = \begin{bmatrix} (L_1 + L_2)\sin\alpha - (D_1 + D_2)\cos\alpha \\ 0 \\ -(L_1 + L_2)\cos\alpha - (D_1 + D_2)\sin\alpha \end{bmatrix} \qquad (17)$$

and the moments created by the aerodynamic forces produced by the wings $\tau_w$ are

$$\tau_w = \begin{bmatrix} l_w[(L_2 - L_1)\cos\alpha + (D_2 - D_1)\sin\alpha] \\ l_c[(L_2 + L_1)\cos\alpha + (D_2 + D_1)\sin\alpha] \\ l_w[(D_2 - D_1)\cos\alpha + (L_1 - L_2)\sin\alpha] \end{bmatrix} \qquad (18)$$

The parameters of fuselage lift and drag are presented as follows:

$$L_f = 0.5\rho C_{lf} S_f (\dot{x}_b^2 + \dot{z}_b^2), D_f = 0.5 C_{df} S_f \rho (\dot{x}_b^2 + \dot{z}_b^2)$$

$$C_{lf} = C_{lf\alpha}\alpha, C_{df} = C_{df0} + C_{df\alpha}\alpha$$

$$C_{lf\alpha} = 0.0802, C_{df0} = 0.0063, C_{df\alpha} = 0.0094 \qquad (19)$$

where $L_f$ and $D_f$ are the lift and drag forces generated by the fuselage, respectively; $C_{lf}$ is lift coefficient; $C_{df}$ is the drag coefficient; $C_{df0}$ is the constant in the coefficients of drag force. Then forces on the fuselage $F_f$ in body frame are written as

$$F_f = \begin{bmatrix} L_f \sin\alpha - D_f \cos\alpha \\ 0 \\ -L_f \cos\alpha - D_f \sin\alpha \end{bmatrix} \qquad (20)$$

*2) The parameters of co-axial counter-rotating propellers*

$$v_u = \lambda_c R_d \omega_u, \ v_l = \lambda_c R_d \omega_l, \ \lambda_c = 0.2673 \qquad (21)$$

where $v_u$ and $v_l$ are the induced velocities of upper and lower rotors, respectively; $\omega_u$ and $\omega_l$ are the rotational speeds of the upper and lower rotors, respectively; $\lambda_c$ is a positive non-dimensional quantity, which is called as induced inflow ratio; $R_d$ is the radius of the rotor disk of the co-axial counter rotating propellers.

*3) The parameters of quad rotors and vanes are shown as follows:*

The lift forces of quad rotors 1-4:

$$F_{ri} = b\omega_i^2, (i = 1, 2, 3, 4), b = 5 \times 10^{-4} \qquad (22)$$

where $F_{r1}, \cdots, F_{r4}$ are the thrust forces generated by the four rotors, respectively; $\omega_1, \cdots, \omega_4$ are the angular rates of the rotors, respectively; $b$ is the coefficient of lift force for each rotor. The sum of the quad rotors thrusts can be written as

$$F_r = \begin{bmatrix} \sum_{i=1}^{4} F_{ri} & 0 & 0 \end{bmatrix}^T = \begin{bmatrix} b\sum_{i=1}^{4} \omega_i^2 & 0 & 0 \end{bmatrix}^T \qquad (23)$$

As the blades of quad rotors 1-4 rotate, they are subject to drag forces which produce torques around the aerodynamic center $O_{ri}$. These moments act in opposite direction relative to $\omega_i$. In hover, the reactive torque generated in free air by the rotor due to rotor drag is given by:

$$\tau_{ri} = k\omega_i^2, (i = 1, 2, 3, 4), k = 3 \times 10^{-5} \qquad (24)$$

where $k$ is a positive constant. Because coefficient $k$ is very small, reactive torques can't provide the sufficient torque. A torque amplifier for roll dynamics is designed as follow (see Figure 8):

$$\tau_{ai} = \sqrt{2} f_{ai} l_3 / 2, \ i = 1, \cdots, 4 \qquad (25)$$

$$f_{ai} = k_f \omega_i^2, k_f = c_a \delta_a \qquad (26)$$

where, $\delta_a$ is the deflection angle of vane, $c_a$ is the coefficient of moment for a fairing. From the simulation results of Fluent, we can obtain a conclusion that when each vane has single bade, the vane angle is $0.13686 rad$ (i.e., 7.84deg or so). Therefore, the sum of torques of each rotor with a vane is

$$\tau_{ri} + \tau_{ai} = (k + \sqrt{2}k_f l_3/2)\omega_i^2, \quad i = 1, \cdots, 4 \tag{27}$$

Therefore, the moments created by the rotors are

$$\tau_r = \begin{bmatrix} \sum_{i=1}^{4}(-1)^{i+1}(\tau_{ri} + \tau_{ai}) \\ [(F_{r1} + F_{r2}) - (F_{r3} + F_{r4})]\frac{l_3}{2} \\ [(F_{r2} + F_{r3}) - (F_{r1} + F_{r4})]\frac{l_3}{2} \end{bmatrix} \tag{28}$$

and the gyroscopic effects of the propellers $\tau_{gyro}$ can be written as

$$\tau_{gyro} = \sum_{i=1}^{4} J_r \Omega_\Lambda \times e_z (-1)^i \omega_i \tag{29}$$

where $J_r$ is the moment of inertia of each rotor.

### 3.4 Aerodynamic analysis of co-axial counter-rotating propellers 5 and 6 with quad rotors 1-4

The side slip angle $\beta$ is assumed to be low during mode transitions. The performance treatment of co-axial counter-rotating propellers 5-6 and quad rotors 1-4 in mode transition is shown in Figure 11.

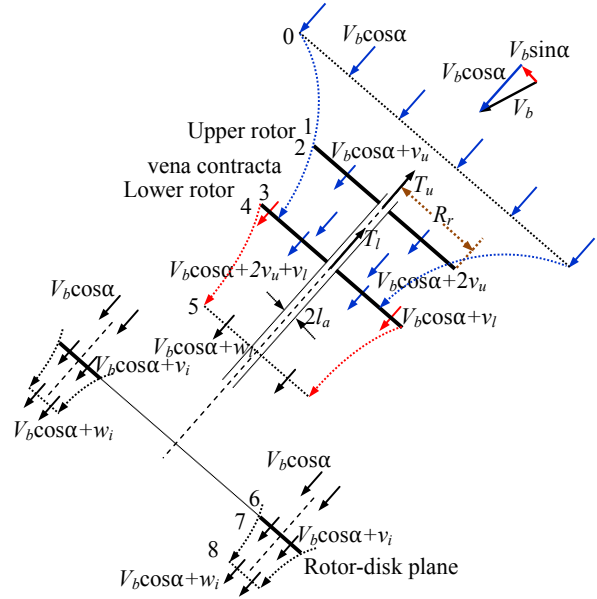

Figure 11 Flow model analysis, where the lower rotor is considered to operate in the fully developed slipstream of the upper rotor

One advantage of the co-axial counter-rotating propeller design is that the net size of the rotor(s) is reduced (for the aircraft gross weight) because each rotor now provides thrust. In addition, no additional rotor is required for anti-torque purposes, so that all power can be devoted to providing useful thrust and performance. However, the rotors and their wakes interact with one another, producing a somewhat more complicated flow field than is found with a single rotor, and this interacting flow incurs a loss of net rotor system aerodynamic efficiency.

The main reason for the over-prediction of induced power is related to the actual (finite) spacing between the rotors. Generally, on co-axial designs the rotors are spaced sufficiently far that the lower rotor operates in the vena contracta of the upper rotor (the radius length of a propeller). Based on ideal flow considerations, this means that only half of the area of the lower operates in an effective velocity induced by the upper rotor.

In [7], the aerodynamic analysis was given when each propeller provides an equal fraction of the total system thrust in hover. However, the undesired torque exists during mode transition. In the following, we will give the aerodynamic analysis when the co-axial is operated at equal rotor torque (power) during mode transition.

We assume that the performance of the upper rotor is not influenced by the lower rotor. Let $V_b$ be the relative velocity far upstream relative to the rotor. The vena contracta of the upper rotor is an area of $A/2$ with velocity $2v_u+V_b\cos\alpha$. Therefore, at the plane of the lower rotor there is a velocity of $2v_u+v_l+V_b\cos\alpha$ over the inner one-half of the disk area (See Figure 11).

**Theorem 1:** For co-axial counter-rotating propellers 5-6, when the co-axial is operated at equal rotor torque (power), there exist the bounded functions $\Gamma_\omega(\alpha,V_b)$, $\Gamma_{uv}(\alpha,V_b)$ and $\Gamma_{Fc}(\alpha,V_b)$, such that the following relations hold:

$$\omega_l = k_{uv}\omega_u + \Gamma_\omega(\alpha,V_b), \quad v_l = k_{uv}v_u + \Gamma_{uv}(\alpha,V_b) \tag{30}$$

and

$$F_c = F_{cu} + F_{cl} = \bar{k}_u\omega_u^2 + \Gamma_{Fc}(\alpha,V_b) \tag{31}$$

where, $k_{uv} = 0.4376$, $\bar{k}_u = 3.3913\rho A\lambda_c^2 R_d^2$; $\Gamma_\omega(\alpha,V_b)$, $\Gamma_{uv}(\alpha,V_b)$ and $\Gamma_{Fc}(\alpha,V_b)$ are the bounded functions of $\alpha$ and $V_b$, and $\Gamma_\omega(0,0)=0$, $\Gamma_{uv}(0,0)=0$ and $\Gamma_{Fc}(0,0)=0$; $F_{cu}$ and $F_{cl}$ are the thrusts on the upper and lower rotors, respectively; $\omega_u$ and $\omega_l$ are the rotational speeds of the upper and lower rotors, respectively; $v_u$ and $v_l$ are the induced velocities of the upper and lower rotors, respectively; $\lambda_c$ is a positive non-dimensional quantity, which is called as induced inflow ratio, such that $v_u = \lambda_c R_d \omega_u$ holds; $R_d$ is the radius of the rotor disk of one of the co-axial counter-rotating propellers.

The proof of Theorem 1 is presented in Appendix.

**Remark 1:** Introducing quad rotors 1-4 increases the sum of induced power factors with respect to that of only the co-axial counter-rotating propellers. In hover, the induced power factor for all the rotors (co-axial counter-rotating propellers 5-6 and quad rotors 1-4) is given by

$$\kappa_{inth} < 1.2809 \tag{32}$$

Therefore, the use of quad rotors 1-4 increases the efficiency of thrusts. In fact,

*1) The induced power factor of co-axial counter-rotating propeller*

In hover, when the rotors are operating in isolation, we obtain the thrust of each rotor as follow:

$$F_{ce} = F_c/2 = 1.6957\rho A v_u^2 \tag{33}$$

Furthermore, we obtain

$$F_{ce} = 2\rho A v_e^2 \tag{34}$$

where $v_e$ is the induced velocity for each rotor operating in isolation. Accordingly, it can be followed that

$$v_e = 0.9208 v_u \tag{35}$$

The power for each rotor in isolation can be written as

$$P_{ce} = F_{ce}v_e = 2\rho A v_e^3 = 1.5614\rho A v_u^3 \tag{36}$$

Therefore, when the co-axial propellers are operated at equal torque (power), the induced power factor is given by

$$\kappa_{int} = (2P_c)/(2P_{ce}) = 2\rho A v_u^3 / (1.5614\rho A v_u^3) = 1.2809 \tag{37}$$

*2) The induced power factor of co-axial counter-rotating propeller with quad rotors 1-4*

Let the thrust of one of quad rotors be

$$F_{ri} = \varsigma F_c \tag{38}$$

where $\varsigma \in (0,1)$ is a positive constant. Therefore,

$$F_{ri} = 3.3913\varsigma\rho A v_u^2 = 2\rho A_q v_i^2 \tag{39}$$

where $A_q$ is the disk area of the rotor. Then,

$$v_i = 1.3022 v_u \sqrt{\varsigma A / A_q} \tag{40}$$

The power for each rotor is

$$P_{ri} = 2\rho A_q v_i^3 = 4.4164\rho(\varsigma A)^{3/2} v_u^3 / (A_q)^{1/2} = 4.4164\rho A \varsigma^{3/2} \eta^{1/2} v_u^3 \tag{41}$$

where $\eta = A/A_q$. There, from (113), (36), (41), and $P_{cu} = F_{cu}v_u = 2\rho A v_u^3$, induced power factor for all the propellers is given by

$$\kappa_{inth} = \frac{2P_{cu} + 4P_{ri}}{2P_{ce} + 4P_{ri}} = \frac{2 + 2\times 4.4164\varsigma^{3/2}\eta^{1/2}}{1.5614 + 2\times 4.4164\varsigma^{3/2}\eta^{1/2}} \tag{42}$$

It is found that

$$\kappa_{inth} < 1.2809 \tag{43}$$

Therefore, the use of quad rotors 1-4 increases the efficiency of thrusts.

### 3.5 Measurement sensors and actuators

*Position* (*X,Y*) and *Velocity* ($\dot{X}, \dot{Y}$) can be obtained by Global Positioning System (GPS). The position data from GPS is sent to the processor in the aircraft for feedback control. *Altitude Z* and its velocity $\dot{Z}$ are measured by an altimeter. *The relative wind speed* ($\dot{x}_b, \dot{y}_b, \dot{z}_b$) is measured by the airspeed tube. *Attitude* $(\psi,\theta,\phi)^T$ and *attitude rate* $(\dot{\psi},\dot{\theta},\dot{\phi})^T$ can be obtained by an Inertial Measurement Unit (IMU). Most common attitude sensors are based on gyros. *Angle of attack* is measured by an angle of attack sensor. Angle of attack is quite simply the angle between the wing chord and the oncoming air that the wing is flying through. The uncertain force $F_d$ and moment $\tau_d$ should be reconstructed from the known information. The thrust $F_c$ generated by the co-axial counter-rotating propellers and rotor thrusts $F_{r1}$, $F_{r2}$, $F_{r3}$ $F_{r4}$, are selected as the control actuators.

### 4  OBSERVER DESIGN

From Eqs. (9) and (10), in systems (2) and (3), $F_d$ and $\tau_d$ are the unknown external disturbances in the position and attitude dynamics, respectively. Moreover, for the co-axial counter-rotating propellers, the uncertain nonlinear terms exist in the relationship between the thrust and the rotational speed. In order to reconstruct these unknown terms, we will design the finite-time convergent observers.

*4. 1 Finite-time convergent observer*

Considering (9), (10) and (31), the systems (2) and (3) can be rewritten as

$$\dot{\upsilon}_\Gamma = \frac{1}{m}R(F_{c0} + F_r + F_w + F_f) - gE_z + \frac{1}{m}R\overline{F}_d \tag{44}$$

$$\dot{\Omega}_\Lambda = -J^{-1}\Omega_\Lambda \times J\Omega_\Lambda + J^{-1}(\tau_r + \tau_w + \tau_{gyro}) + J^{-1}\tau_d \tag{45}$$

where $F_{c0} = \overline{k}_u \omega_u^2$ and $\overline{F}_d = F_d + \Gamma_{Fc}(\alpha, V_b)$.

In systems (44) and (45), $\overline{F}_d$ and $\tau_d$ are uncertain vectors. A general form for systems (44) and (45) can be described as

$$\dot{\zeta}_{i1} = \Xi_i + \delta_{d,i} \tag{46}$$

where $i=1, \cdots, 6$, and

$$\upsilon_\Gamma = [\zeta_{11} \ \zeta_{21} \ \zeta_{31}]^T, \ \Omega_\Lambda = [\zeta_{41} \ \zeta_{51} \ \zeta_{61}]^T; \ \frac{1}{m}R(F_{c0} + F_r + F_w + F_f) - gE_z = [\Xi_1 \ \Xi_2 \ \Xi_3]^T$$

$$-J^{-1}\Omega_\Lambda \times J\Omega_\Lambda + J^{-1}(\tau_r + \tau_w + \tau_{gyro}) = [\Xi_4 \ \Xi_5 \ \Xi_6]^T; \ \frac{1}{m}R\overline{F}_d = [\delta_{d,1} \ \delta_{d,2} \ \delta_{d,3}]^T, \ J^{-1}\tau_d = [\delta_{d,4} \ \delta_{d,5} \ \delta_{d,6}]^T \tag{47}$$

**Assumption 2:** Uncertainties $\delta_{d,i}$, $i=1,\cdots,6$, have the following dynamics:

$$\dot{\delta}_{d,i} = \eta_i(t), \ i=1,\cdots,6 \tag{48}$$

where $\eta_i(t)$ is bounded, and $\sup_{t\in[0,\infty)}|\eta_i(t)| \leq L_{di}$, $L_{di}$ is a positive constant, $i=1,\cdots,6$. In fact, this assumption is satisfied with almost all engineering applications, for instance, the dynamics of crosswind and the uncertainties in the aircraft.

Let $\zeta_{i2} = \delta_{d,i}$ and $\dot{\zeta}_{i2} = \eta_i(t)$ (where $i=1,\cdots,6$), Eq. (46) can be extended to

$$\begin{aligned}\dot{\zeta}_{i1} &= \zeta_{i2} + \Xi_i \\ \dot{\zeta}_{i2} &= \eta_i(t) \\ y_{opi} &= \zeta_{i1}\end{aligned} \tag{49}$$

where $[y_{op1} \ y_{op2} \ y_{op3} \ y_{op4} \ y_{op5} \ y_{op6}]^T = [\upsilon_\Gamma^T \ \Omega_\Lambda^T]^T$. $\upsilon_\Gamma$ is the aircraft velocity relative to the inertial frame $\Gamma$, and $\Omega_\Lambda$ is the angular velocity of the airframe expressed in body frame $\Lambda$. They are defined in Eqs. (2) and (3). In order to reconstruct the uncertainties in systems (44) and (45) (i.e., (49)), we present the finite-time convergent observers, and a theorem is given as follow.

**Theorem 2:** For system (49), the finite-time convergent observers are designed as

$$\begin{aligned}\dot{\hat{\zeta}}_{i1} &= \hat{\zeta}_{i2} + \Xi_i - k_{i,1}|\hat{\zeta}_{i1} - \zeta_{i1}|^{1/2}\operatorname{sign}(\hat{\zeta}_{i1} - \zeta_{i1}) \\ \dot{\hat{\zeta}}_{i2} &= -k_{i,2}\operatorname{sign}(\hat{\zeta}_{i1} - \zeta_{i1})\end{aligned} \tag{50}$$

where $k_{i,1} > 0$ and $k_{i,2} > 0$; $i=1,\cdots,6$; $\hat{\upsilon}_\Gamma = [\hat{\zeta}_{11} \ \hat{\zeta}_{21} \ \hat{\zeta}_{31}]^T$, $\hat{\Omega}_\Lambda = [\hat{\zeta}_{41} \ \hat{\zeta}_{51} \ \hat{\zeta}_{61}]^T$; $\frac{1}{m}R\hat{\overline{F}}_d = [\hat{\zeta}_{12} \ \hat{\zeta}_{22} \ \hat{\zeta}_{32}]^T$, $J^{-1}\hat{\tau}_d = [\hat{\zeta}_{42} \ \hat{\zeta}_{52} \ \hat{\zeta}_{62}]^T$; and $k_{i,2} > |\eta_i(t)|$. Then, a finite time $t_s > 0$ exists, for $t \geq t_s$, such that the following statements hold:

$$\hat{\upsilon}_\Gamma = \upsilon_\Gamma, \ \hat{\Omega}_\Lambda = \Omega_\Lambda, \ \frac{1}{m}R\hat{\overline{F}}_d = \frac{1}{m}R\overline{F}_d, \ J^{-1}\hat{\tau}_d = J^{-1}\tau_d \tag{51}$$

The proof of Theorem 2 is presented in Appendix.

*4.2 Robustness analysis in frequency domain for the finite-time convergent observer*

In a practical problem, high-frequency noise exists in measurement output. The following analysis concerns the robustness behavior of the presented observer under high-frequency noise.

For the presented nonlinear observer, an extended version of the frequency response method, describing function method [36, 37], can be used to approximately analyze and predict the nonlinear behaviors of the observer. Even though it is only an approximation method, the desirable properties it inherits from the frequency response method, and the shortage of other, systematic tools for nonlinear observer analysis, make it an indispensable component of the bag of tools of practicing control engineers. By describing function method, it can be found that the presented observer leads to perform rejection of high-frequency noise.

For system (49), let $\xi_{i1} = \zeta_{i1}, \xi_{i2} = \zeta_{i2} + \Xi_i$, and $\dot{\Xi}_i = \sigma_i(t)$, then $\dot{\xi}_{i2} = \dot{\zeta}_{i2} + \dot{\Xi}_i = \eta_i(t) + \sigma_i(t)$. We know that $\sup_{t \in [0,\infty)} |\eta_i(t)| \le L_d$ and $\sup_{t \in [0,\infty)} |\sigma_i(t)| \le H_d$. Then, $\sup_{t \in [0,\infty)} |\eta_i(t) + \sigma_i(t)| \le L_d + H_d$. Therefore, system (49) can be rewritten as

$$\dot{\xi}_{i1} = \xi_{i2}$$
$$\dot{\xi}_{i2} = \eta_i(t) + \sigma_i(t) \tag{52}$$
$$y_{opi} = \xi_{i1}$$

Accordingly, for system (52), the observer (50) can be transferred to

$$\dot{\hat{\xi}}_{i1} = \hat{\xi}_{i2} - k_{i,1} |\hat{\xi}_{i1} - \xi_{i1}|^{1/2} \operatorname{sign}(\hat{\xi}_{i1} - \xi_{i1})$$
$$\dot{\hat{\xi}}_{i2} = -k_{i,2} \operatorname{sign}(\hat{\xi}_{i1} - \xi_{i1}) \tag{53}$$

where $\hat{\xi}_{i1}$ tracks the output $y_{opi} = \xi_{i1}$; $\hat{\xi}_{i2}$ estimates the first-order derivative of $y_{opi} = \xi_{i1}$, i.e., the state $\xi_{i2}$. The frequency characteristic of (53) is analyzed as follow.

Let $\hat{\xi}_{i1} - y_{opi} = A_0 \sin(\omega_0 t)$, where $A_0$ and $\omega_0$ are the magnitude and frequency of $\hat{\xi}_{i1} - y_{opi}$, respectively. For the following nonlinear functions

$$|A_0 \sin(\omega t)|^{1/2} \operatorname{sign}(\sin(\omega_0 t)) \text{ and } \operatorname{sign}(\sin(\omega_0 t)), \tag{54}$$

their describing functions can be obtained, respective, as follows:

$$N_1(A_0) = \frac{2}{A\pi} \int_0^\pi |A_0 \sin(\omega \tau)|^{1/2} \operatorname{sign}(\sin(\omega_0 \tau)) \sin(\omega_0 \tau) d\omega\tau$$
$$= \frac{2}{A_0^{1/2} \pi} \int_0^\pi |\sin(\omega_0 \tau)|^{3/2} d\omega_0 \tau = \frac{\Delta_1}{A_0^{1/2}} \tag{55}$$

$$N_2(A_0) = \frac{2}{A_0 \pi} \int_0^\pi \operatorname{sign}(\sin(\omega_0 \tau)) \sin(\omega_0 \tau) d\omega\tau$$
$$= \frac{2}{A_0 \pi} \int_0^\pi |\sin(\omega_0 \tau)| d\omega_0 \tau = \frac{4}{A_0 \pi} \tag{56}$$

where $\Delta_1 = \frac{2}{\pi} \int_0^\pi |\sin(\omega \tau)|^{3/2} d\omega\tau = 1.1128$. Therefore, the linearization of continuous observer (53) is

$$\dot{\hat{\xi}}_{i1} = \hat{\xi}_{i2} - k_{i,1} N_1(A_0)(\hat{\xi}_{i1} - \xi_{i1})$$
$$\dot{\hat{\xi}}_{i2} = -k_{i,2} N_2(A_0)(\hat{\xi}_{i1} - \xi_{i1}) \tag{57}$$

i.e.,

$$\dot{\hat{\xi}}_{i1} = \hat{\xi}_{i2} - \frac{1.1128 k_{i,1}}{A_0^{1/2}} (\hat{\xi}_{i1} - \xi_{i1})$$

$$\dot{\hat{\xi}}_{i2} = -\frac{4 k_{i,2}}{A_0 \pi} (\hat{\xi}_{i1} - \xi_{i1})$$

(58)

From the Routh-Hurwitz stability criterion, polynomial $s^2 + \frac{1.1128 k_{i,1}}{A_0^{1/2}} s + \frac{4 k_{i,2}}{A_0 \pi}$ is Hurwitz if $k_{i,1} > 0$ and $k_{i,2} > 0$.

For system (58), the following transfer functions are obtained:

i) The transfer function I from $\xi_{i1}$ to $\hat{\xi}_{i1}$ is

$$G_1(s) = \frac{\frac{1.1128 k_{i,1}}{A_0^{1/2}} s + \frac{4 k_{i,2}}{A_0 \pi}}{s^2 + \frac{1.1128 k_{i,1}}{A_0^{1/2}} s + \frac{4 k_{i,2}}{A_0 \pi}}$$

(59)

ii) The transfer function II from $\xi_{i1}$ to $\hat{\xi}_{i2}$ is

$$G_2(s) = \frac{s \frac{4 k_{i,2}}{A_0 \pi}}{s^2 + s \frac{1.1128 k_{i,1}}{A_0^{1/2}} + \frac{4 k_{i,2}}{A_0 \pi}}$$

(60)

The effects of the observer parameters on the robustness are analyzed as follows.

### 4.2.1 Frequency characteristic with the change in $A_0$

From the transfer functions (59), (60), and the conditions of the observer (50), the parameters are selected as follows: $k_{i,1} = 6$, $k_{i,2} = 8$; $A_0 = 10, 1, 0.1, 0.01$, respectively. The Bode plots for the transfer functions are described in Fig. 12a and b, respectively. It is found that when the tracking error magnitude $A_0$ is large, the cutoff frequency is relatively small and much noise is reduced sufficiently and that when the tracking error magnitude $A_0$ is small, the cutoff frequency is relatively large and the signal with higher frequency can be estimated. Therefore, it is confirmed that the presented observer leads to perform rejection of high-frequency noise.

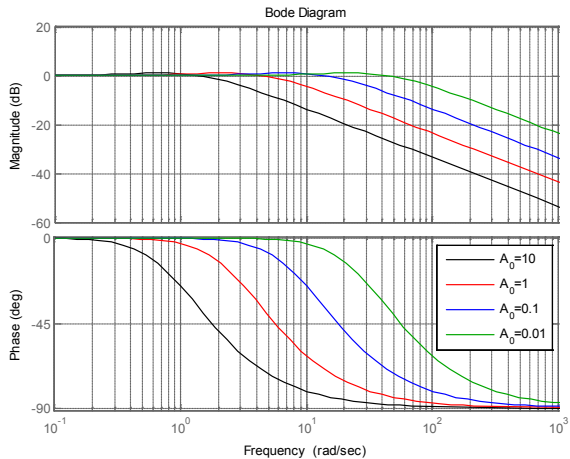
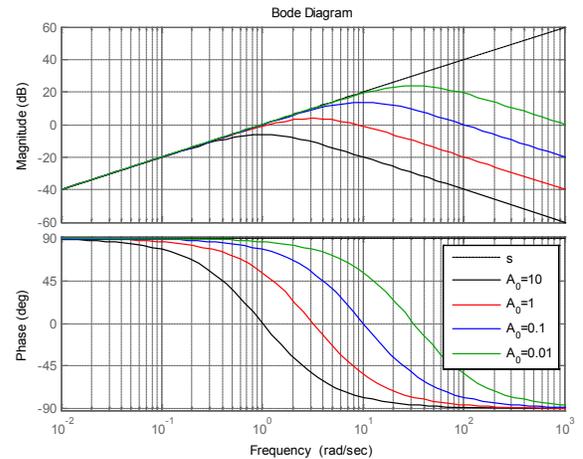

(a) Bode plot of transfer function $G_1(s)$   (b) Bode plot of transfer function $G_2(s)$

Figure 12 Bode plots with the change in $A_0$

# 5 CONTROLLER DESIGN

In this section, a control law is derived for the purpose of attitude stabilization and trajectory tracking. Suppose the reference trajectory and its finite order derivatives are bounded, and can be directly generated.

## 5.1 Controller design for attitude dynamics

From (14), for the desired altitude angle $\Theta_{\Gamma d} = (\phi_d, \theta_d, \psi_d)^T$, the desired attitude in quaternion expression can be obtained as

$$
\begin{aligned}
q_{0d} &= c_{\phi_d/2} c_{\theta_d/2} c_{\psi_d/2} + s_{\phi_d/2} s_{\theta_d/2} s_{\psi_d/2} \\
q_{1d} &= s_{\phi_d/2} c_{\theta_d/2} c_{\psi_d/2} + c_{\phi_d/2} s_{\theta_d/2} s_{\psi_d/2} \\
q_{2d} &= c_{\phi_d/2} s_{\theta_d/2} c_{\psi_d/2} + s_{\phi_d/2} c_{\theta_d/2} s_{\psi_d/2} \\
q_{3d} &= c_{\phi_d/2} c_{\theta_d/2} s_{\psi_d/2} + s_{\phi_d/2} s_{\theta_d/2} c_{\psi_d/2}
\end{aligned}
\quad (61)
$$

Moreover, from (7) and (8), we can obtain

$$
\Omega_{\Lambda d} = \begin{bmatrix} p_d \\ q_d \\ r_d \end{bmatrix} = \mathbb{Z}_d \Omega_{\Gamma d} = \mathbb{Z}_d \begin{bmatrix} \dot{\phi}_d \\ \dot{\theta}_d \\ \dot{\psi}_d \end{bmatrix}
\quad (62)
$$

where

$$
\mathbb{Z}_d = \begin{bmatrix} 1 & 0 & -s_{\theta_d} \\ 0 & c_{\phi_d} & s_{\phi_d} c_{\theta_d} \\ 0 & -s_{\phi_d} & c_{\phi_d} c_{\theta_d} \end{bmatrix}
\quad (63)
$$

The angular velocity denoted by $\Omega_\Lambda$ can be computed from (13) as follow

$$\Omega_\Lambda = 2(q_0 \dot{q} - \dot{q} q_0) - 2S(q)\dot{q} \quad (64)$$

For the desired unit quaternion $Q_d = \{q_{0d}, q_d\} \in R \times R^3$ that is constructed to satisfy

$$q_d^T q_d + q_{0d}^2 = 1 \quad (65)$$

The desired quaternion is related to the desired angular velocity denoted by $\Omega_{\Lambda d} \in R^3$, through the following dynamic equation

$$\dot{q}_d = \frac{1}{2}(S(q_d) + q_{0d} I)\Omega_{\Lambda d} \quad (66)$$

$$\dot{q}_{0d} = -\frac{1}{2} q_d^T \Omega_{\Lambda d} \quad (67)$$

It is noted that Eqs. (66) and (67) can be used to explicitly compute an expression for $\Omega_{\Lambda d}$ as shown bellow

$$\Omega_{\Lambda d} = 2(q_{0d} \dot{q}_d - \dot{q}_d q_{0d}) - 2S(q_d)\dot{q}_d \quad (68)$$

To quantify the mismatch between the actual and desired attitude, the quaternion tracking error $\bar{e} \triangleq \{e_0, e\} \in R \times R^3$ is defined as shown below [38]

$$e_0 = q_0 q_{0d} + q^T q_d \quad (69)$$

$$e = q_{0d} q - q_0 q_d + S(q) q_d \quad (70)$$

And the tracking angular velocity error is defined as follow

$$\tilde{\Omega}_\Lambda = \Omega_\Lambda - \Omega_{\Lambda d} \tag{71}$$

From Eqs. (13), (45) and (66), (67), (69) and (70), we obtain the attitude error dynamics:

$$\dot{\tilde{\Omega}}_\Lambda = -J^{-1}(\tilde{\Omega}_\Lambda + \Omega_{\Lambda d}) \times J(\tilde{\Omega}_\Lambda + \Omega_{\Lambda d}) + J^{-1}\tau_r + J^{-1}(\tau_w + \tau_{gyro}) + J^{-1}\tau_d - \dot{\Omega}_{\Lambda d} \tag{72}$$

$$\dot{e} = \frac{1}{2}(S(e) + e_0 I_3)\tilde{\Omega}_\Lambda \tag{73}$$

$$\dot{e}_0 = -\frac{1}{2}e^T \tilde{\Omega}_\Lambda \tag{74}$$

Let $M_q = S(e) + e_0 I_3$, therefore,

$$\dot{e} = \frac{1}{2}M_q e_{\Omega\Lambda} \tag{75}$$

**Theorem 3:** For the attitude error dynamics (72), (73) and (74), if the controller is designed as

$$\tau_r = (\tilde{\Omega}_\Lambda + \Omega_{\Lambda d}) \times J(\tilde{\Omega}_\Lambda + \Omega_{\Lambda d}) - (\tau_w + \tau_{gyro}) - \hat{\tau}_d + J\dot{\Omega}_{\Lambda d} - 2JM_q^{-1}(k_{a1}e + k_{a2}\dot{e}) - JM_q^{-1}\dot{M}_q \tilde{\Omega}_\Lambda \tag{76}$$

where $k_{a1}, k_{a2} > 0$, then the attitude error dynamics (72), (73) and (74) rendering by controller (76) will converge asymptotically to the origin, i.e., $e \to 0$ and $\dot{e} \to 0$ as $t \to \infty$.

The proof of Theorem 3 is presented in Appendix.

*5.2 Controller design for position dynamics*

For the position dynamics (44), let the reference trajectory vector be $P_{\Gamma d} = (x_d, y_d, z_d)^T$, then the system error for the position dynamics can be written as

$$\begin{aligned} \dot{e}_{\Gamma 1} &= e_{\Gamma 2} \\ \dot{e}_{\Gamma 2} &= \frac{1}{m}[F_p + R(F_w + F_f) + R\overline{F}_d] - gE_z - \ddot{P}_{\Gamma d} \end{aligned} \tag{77}$$

where $e_{\Gamma 1} = P_\Gamma - P_{\Gamma d}$, $e_{\Gamma 2} = \upsilon_\Gamma - \dot{P}_{\Gamma d}$, and

$$F_p = R(F_{c0} + F_r), \quad F_{c0} = \overline{k}_u \omega_u^2, \quad \overline{F}_d = F_d + \Gamma_{Fc}(\alpha, V_b) \tag{78}$$

**Theorem 4:** For the system error dynamics (77), if the controller is designed as

$$F_p = -R\hat{\overline{F}}_d - R(F_w + F_f) + mgE_z + m\ddot{P}_{\Gamma d} - k_{p1}me_{\Gamma 1} - k_{p2}me_{\Gamma 2} \tag{79}$$

where $k_{p1}, k_{p2} > 0$, then the position error dynamic system (77) rendering by controller (79) will converge asymptotically to the origin, i.e., the tracking error $e_{\Gamma 1} \to 0$ and $e_{\Gamma 2} \to 0$ as $t \to \infty$.

The proof of Theorem 4 is presented in Appendix.

*5.3 The implement of autopilot*

From (24) and (28), we known that

$$\tau_r = \begin{bmatrix} \tau_{r1} \\ \tau_{r2} \\ \tau_{r3} \end{bmatrix} = \begin{bmatrix} \sum_{i=1}^{4}(-1)^{i+1}\left(k+\sqrt{2}\,l_3 k_f/2\right)\omega_i^2 \\ \left(\omega_1^2+\omega_2^2-\omega_3^2-\omega_4^2\right)\dfrac{bl_3}{2} \\ \left(-\omega_1^2+\omega_2^2+\omega_3^2-\omega_4^2\right)\dfrac{bl_3}{2} \end{bmatrix} \quad (80)$$

And from (1), (23), (78) and (79), we can obtain

$$F_{c0} + F_r = \bar{k}_u \omega_u^2 + b\sum_{i=1}^{4}\omega_i^2 = \|F_p\|_2 = \|R(F_{c0}+F_r)\|_2 \quad (81)$$

We allocate $\bar{k}_u \omega_u^2$ and $b\sum_{i=1}^{4}\omega_i^2$ according to the following relation:

$$\bar{k}_u \omega_u^2 = Kb\sum_{i=1}^{4}\omega_i^2 \quad (82)$$

where $K > 1$ is decided by the maneuverability requirement of the desired trajectory. From Equations (55), (65), (66), (68) and (69), we can carry out $\omega_1, \omega_2, \omega_3, \omega_4$ and $\omega_u$. In fact,

$$\begin{aligned}
\omega_1^2 + \omega_2^2 + \omega_3^2 + \omega_4^2 &= \frac{\|F_p\|_2}{b(1+K)} \\
\omega_1^2 - \omega_2^2 + \omega_3^2 - \omega_4^2 &= \frac{\tau_{r1}}{k+\sqrt{2}\,l_3 k_f/2} \\
\omega_1^2 + \omega_2^2 - \omega_3^2 - \omega_4^2 &= \frac{2\tau_{r2}}{bl_3} \\
-\omega_1^2 + \omega_2^2 + \omega_3^2 - \omega_4^2 &= \frac{2\tau_{r3}}{bl_3}
\end{aligned} \quad (83)$$

Therefore, we obtain

$$\omega_1 = \frac{1}{2}\sqrt{\frac{\|F_p\|_2}{b(1+K)} + \frac{\tau_{r1}}{k+\sqrt{2}\,l_3 k_f/2} + \frac{2\tau_{r2}}{bl_3} - \frac{2\tau_{r3}}{bl_3}},\ \omega_2 = \frac{1}{2}\sqrt{\frac{\|F_p\|_2}{b(1+K)} - \frac{\tau_{r1}}{k+\sqrt{2}\,l_3 k_f/2} + \frac{2\tau_{r2}}{bl_3} + \frac{2\tau_{r3}}{bl_3}}$$

$$\omega_3 = \frac{1}{2}\sqrt{\frac{\|F_p\|_2}{b(1+K)} + \frac{\tau_{r1}}{k+\sqrt{2}\,l_3 k_f/2} - \frac{2\tau_{r2}}{bl_3} + \frac{2\tau_{r3}}{bl_3}},\ \omega_4 = \frac{1}{2}\sqrt{\frac{\|F_p\|_2}{b(1+K)} - \frac{\tau_{r1}}{k+\sqrt{2}\,l_3 k_f/2} - \frac{2\tau_{r2}}{bl_3} - \frac{2\tau_{r3}}{bl_3}} \quad (84)$$

From (82), it follows that

$$\omega_u = \sqrt{\frac{K}{\bar{k}_u}\frac{\|F_p\|_2}{1+K}} \quad (85)$$

Therefore, from (30), we obtain

$$\omega_l = k_{uv}\sqrt{\frac{K}{\bar{k}_u}\frac{\|F_p\|_2}{1+K}} \quad (86)$$

**Remark 2:** *Controller design in forward flight*

The aircraft is equipped with ailerons on each half wing. The ailerons are used to control the roll dynamics in forward flight. It could help save energy on the tail rotors.

In forward flight, the control torque is taken as

$$\tau_r = \begin{bmatrix} \tau_{r1} \\ \tau_{r2} \\ \tau_{r3} \end{bmatrix} = \begin{bmatrix} \cos\alpha(L_2 - L_1)l_w \\ (\omega_1^2 + \omega_2^2 - \omega_3^2 - \omega_4^2)\dfrac{bl_3}{2} \\ (-\omega_1^2 + \omega_2^2 + \omega_3^2 - \omega_4^2)\dfrac{bl_3}{2} \end{bmatrix} \tag{87}$$

The deflexion angles of vanes in the torque amplifier are fixed to be zero, and the reactive torques generated in free air by the quad rotors due to rotor drags are restrained by

$$k\sum_{i=1}^{4}(-1)^{i+1}\omega_i^2 = 0 \tag{88}$$

The moments

$$\tau_w = \begin{bmatrix} (D_2 - D_1)\sin\alpha \\ l_c[(L_2 + L_1)\cos\alpha + (D_2 + D_1)\sin\alpha] \\ l_w[(D_2 - D_1)\cos\alpha + (L_1 - L_2)\sin\alpha] \end{bmatrix} \tag{89}$$

generated by fixed wing is taken as a part of uncertainties. Therefore, the moments due to the external disturbances $\bar{\tau}_d$ can be written as

$$\bar{\tau}_d = \tau_w + \tau_d \tag{90}$$

Thus, the total moment $\tau$ become

$$\tau = \tau_r + \tau_{gyro} + \bar{\tau}_d \tag{91}$$

Therefore, from (76), the controller for attitude dynamics in forward flight can be written as

$$\tau_r = (\tilde{\Omega}_\Lambda + \Omega_{\Lambda d}) \times J(\tilde{\Omega}_\Lambda + \Omega_{\Lambda d}) - \tau_{gyro} - \bar{\hat{\tau}}_d + J\dot{\Omega}_{\Lambda d} - 2JM_q^{-1}(k_{a1}e + k_{a2}\dot{e}) - JM_q^{-1}\dot{M}_q\tilde{\Omega}_\Lambda \tag{92}$$

From the controllers (79) and (92), we obtain the following relations

$$\begin{aligned} \omega_1^2 + \omega_2^2 + \omega_3^2 + \omega_4^2 &= \dfrac{\|F_p\|_2}{b(1+K)} \\ \omega_1^2 - \omega_2^2 + \omega_3^2 - \omega_4^2 &= 0 \\ \omega_1^2 + \omega_2^2 - \omega_3^2 - \omega_4^2 &= \dfrac{2\tau_{r2}}{bl_3} \\ -\omega_1^2 + \omega_2^2 + \omega_3^2 - \omega_4^2 &= \dfrac{2\tau_{r3}}{bl_3} \end{aligned} \tag{93}$$

Therefore, we obtain

$$\omega_1 = \dfrac{1}{2}\sqrt{\dfrac{\|F_p\|_2}{b(1+K)} + \dfrac{2\tau_{r2}}{bl_3} - \dfrac{2\tau_{r3}}{bl_3}}, \quad \omega_2 = \dfrac{1}{2}\sqrt{\dfrac{\|F_p\|_2}{b(1+K)} + \dfrac{2\tau_{r2}}{bl_3} + \dfrac{2\tau_{r3}}{bl_3}}$$

$$\omega_3 = \dfrac{1}{2}\sqrt{\dfrac{\|F_p\|_2}{b(1+K)} - \dfrac{2\tau_{r2}}{bl_3} + \dfrac{2\tau_{r3}}{bl_3}}, \quad \omega_4 = \dfrac{1}{2}\sqrt{\dfrac{\|F_p\|_2}{b(1+K)} - \dfrac{2\tau_{r2}}{bl_3} - \dfrac{2\tau_{r3}}{bl_3}} \tag{94}$$

For the fixed wing, from (16), it follows that

$$\begin{aligned} L_i &= 0.5S\rho(\dot{x}_b^2 + \dot{z}_b^2)(C_{L0} + C_{L\alpha}\alpha + C_{L\delta_i}\delta_i) \\ &= 0.5S\rho(\dot{x}_b^2 + \dot{z}_b^2)(C_{L0} + C_{L\alpha}\alpha) + 0.5S\rho(\dot{x}_b^2 + \dot{z}_b^2)C_{L\delta_i}\delta_i \end{aligned} \tag{95}$$

where $i =$, 1, 2. Therefore

$$\cos\alpha(L_2 - L_1)l_w = 0.5l_w S\rho(\dot{x}_b^2 + \dot{z}_b^2)\cos\alpha(C_{L\delta_2}\delta_2 - C_{L\delta_1}\delta_1) \tag{96}$$

Selecting $C_{L\delta_2} = C_{L\delta_2} = C_{L\delta_{1,2}}$, $\delta_1 = -\delta_2 = \delta_{1,2}$, we obtain

$$\cos\alpha(L_2 - L_1)l_w = l_w C_{L\delta_{1,2}} S\rho(\dot{x}_b^2 + \dot{z}_b^2)\delta_{1,2}\cos\alpha \tag{97}$$

From (87) and (97), it follows that

$$\tau_{r1} = \cos\alpha(L_2 - L_1)l_w = l_w C_{L\delta_{1,2}} S\rho(\dot{x}_b^2 + \dot{z}_b^2)\delta_{1,2}\cos\alpha$$

Therefore, the ailerons control law in forward flight can be obtain as follow

$$\delta_{1,2} = \frac{\tau_{r1}}{l_w C_{L\delta_{1,2}} S\rho(\dot{x}_b^2 + \dot{z}_b^2)\cos\alpha} \tag{98}$$

## 6 DESIRED TRAJECTORY AND ATTITUDE ANGLE DURING MODE TRANSITION

Trajectory design is treated independently for both dimensions. For the desired trajectory, let $x_d$, $\dot{x}_d$ and $\ddot{x}_d$ denote the position, velocity and acceleration, respectively, in the level direction; $z_d$, $\dot{z}_d$ and $\ddot{z}_d$ denote the position, velocity and acceleration, respectively, in the vertical direction; The $x_d$ trajectory is velocity based. For a hover-to-level transition, the tail-sitter's velocity will initially be zero and will need to increase to $v_f$ when it is in level flight. For a level-to-hover transition, the velocity will initially be $v_f$ and will then go to zero as the tail-sitter assumes a hover position.

1) For a hover-to-level transition, the trajectory in the $x_d$ direction is given by

$$\ddot{x}_d = \begin{cases} a, t \leq t_m \\ 0, \text{otherwise} \end{cases}, \quad \dot{x}_d = \begin{cases} at + v_0, t \leq t_m \\ v_f = at_m + v_0, \text{otherwise} \end{cases},$$

$$x_d = \begin{cases} 0.5at^2 + v_0 t, t \leq t_m \\ v_f(t - t_m) + 0.5at_m^2 + v_0 t_m, \text{otherwise} \end{cases} \tag{99}$$

where

$$t_m = 2(v_f - v_0)/a \tag{100}$$

And the desired trajectories in the $z_d$ direction is given by

$$z_d = h_0\left(1 - e^{-k_m(0.5at^2 + v_0 t)}\right), \quad \dot{z}_d = h_0 k_m (at + v_0) e^{-k_m(0.5at^2 + v_0 t)}, \quad \ddot{z}_d = h_0 k_m \left(a + k_m (at + v_0)^2\right) e^{-k_m(0.5at^2 + v_0 t)} \tag{101}$$

We can find that $z_d \to h_0, \dot{z}_d \to 0, \ddot{z}_d \to 0$ as $t \to \infty$. Furthermore, we can obtain

$$z_d = h_0\left(1 - e^{-k_m x_d}\right) \tag{102}$$

This space motion trajectory is easy to be implemented. For Eqs. (99)-(101), the parameters are selected as $a = 5m/s^2, v_f = 50m/s, v_0 = 0m/s, h_0 = 30m, k_m = 0.05$. The trajectory for a hover-to-level transition is shown in Figure 13.

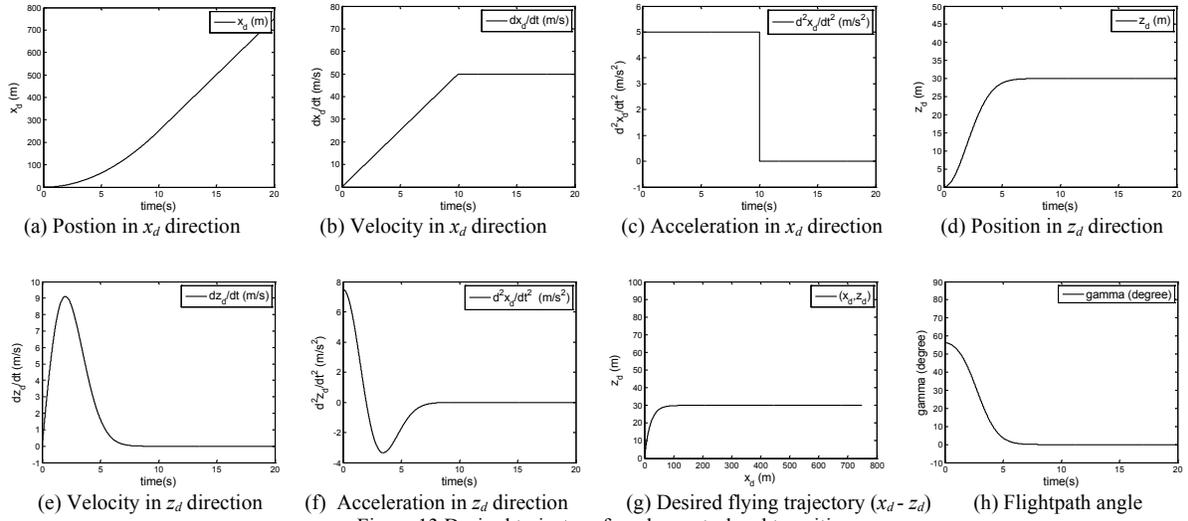

(a) Postion in $x_d$ direction  (b) Velocity in $x_d$ direction  (c) Acceleration in $x_d$ direction  (d) Position in $z_d$ direction
(e) Velocity in $z_d$ direction  (f) Acceleration in $z_d$ direction  (g) Desired flying trajectory ($x_d$ - $z_d$)  (h) Flightpath angle
Figure 13 Desired trajectory for a hover-to-level transition

2) For level-to-hover transition, trajectory in the $x_d$ direction is given by

$$\ddot{x}_d = \begin{cases} -a, t \le t_m \\ 0, \text{otherwise} \end{cases}, \quad \dot{x}_d = \begin{cases} -at + v_f, t \le t_m \\ 0, \text{otherwise} \end{cases}, \quad x_d = \begin{cases} -0.5at^2 + v_f t, t \le t_m \\ -0.5at_m^2 + v_f t_m, \text{otherwise} \end{cases} \tag{103}$$

where

$$t_m = v_f / a \tag{104}$$

And for the desired trajectory in the $z_d$ direction is given by

$$z_d = h_0 \left(1 - e^{-0.5 k_m a t^2}\right), \quad \dot{z}_d = h_0 k_m a t e^{-0.5 k_m a t^2}, \quad \ddot{z}_d = h_0 k_m a \left(1 - k_m a t^2\right) e^{-0.5 k_m a t^2} \tag{105}$$

We can find that $z_d \to h_0, \dot{z}_d \to 0, \ddot{z}_d \to 0$ as $t \to \infty$. For Equations (103)-(105), the parameters are selected as $a = 5m/s^2, v_f = 50m/s, h_0 = 30m, k_m = 0.005$. The trajectory for a level-to-hover transition is shown in Figure 14.

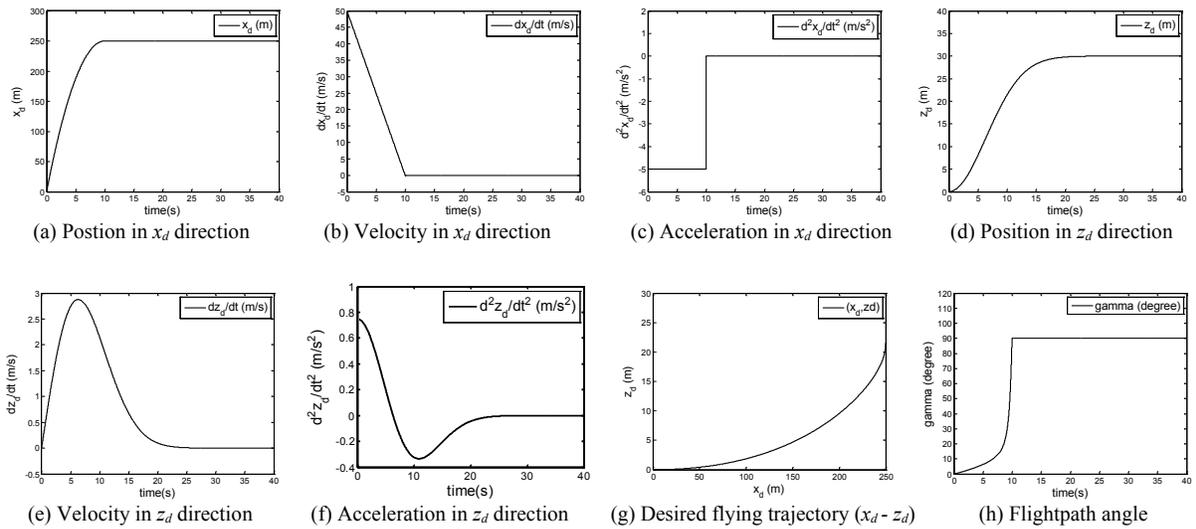

(a) Postion in $x_d$ direction  (b) Velocity in $x_d$ direction  (c) Acceleration in $x_d$ direction  (d) Position in $z_d$ direction
(e) Velocity in $z_d$ direction  (f) Acceleration in $z_d$ direction  (g) Desired flying trajectory ($x_d$ - $z_d$)  (h) Flightpath angle
Figure 14 Desired trajectory for a level-to-hover transition

3) During cruise, let $2L_1 = 2L_2 = mg$. From Equation (16), $(C_{L0} + C_{L\alpha}\alpha_d)S\rho V_f^2 = mg$ is established. Therefore, we obtain the desired angle of attack $\alpha_d = \alpha_{d0}$. For hover-to-level and level-to-hover transitions, we select

$$\alpha_d = \begin{cases} \alpha_{d0}, \gamma \leq \gamma_s \\ 0, \text{ otherwise} \end{cases} \tag{106}$$

where $\gamma_s$ is a large flightpath angle approached to $90°$. During mode transition, the aircraft is required to move in X-Z plane, i.e., $y_d = 0$. For the desired attitude, $(\phi_d, \theta_d, \psi_d) = (0, \theta_d, 0)$. Moreover, for the pitch dynamics, the angle of attack $\alpha$ is required to be kept at a given degree $\alpha_d$. From the desired flying velocity $(\dot{x}_d, \dot{z}_d)$, we can obtain the desired flightpath angle as follow

$$\gamma_d = \arctan^{-1}(\dot{z}_d/\dot{x}_d) \tag{107}$$

Therefore, the desired pitch angle can be given by

$$\theta_d = \alpha_d + \gamma_d \tag{108}$$

## 7 COMPUTATIONAL ANALYSIS AND SIMULATION EXPERIMENTS

The parameters of aircraft model and flight are shown in Table 1, and the simulink of the tail-sitter aircraft control system is described in Figure 15.

For hover-to-level transition, the desired trajectory is shown in Eqs. (99)-(101) and (107) (see Figure 13). The initial attitude angle vector is $(\phi, \theta, \psi) = (0, 90°, 0)$, position vector $(x, y, z) = (0, 0, 0)$, and velocity vector $(\dot{x}, \dot{y}, \dot{z}) = (0, 0, 0)$. The initial values of angular rate of rotor and rolling angle of flap are

$$(\omega_u, \omega_1, \omega_2, \omega_3, \omega_4) = (290, 310.1, 310.1, 310.1, 310.1)$$

For level-to-hover transition, the desired trajectory is shown in Equations (103)-(105) and (107) (see Figure 14), the initial attitude angle vector $(\phi, \theta, \psi) = (0, 5°, 0)$, the position $(x, y, z) = (0, 0, 0)$, the velocity vector $(\dot{x}, \dot{y}, \dot{z}) = (50, 0, 0)$. The initial angular rate of rotor and rolling angle of flap are

$$(\omega_u, \omega_1, \omega_2, \omega_3, \omega_4) = (102, 74.4, 74.4, 74.4, 74.4)$$

The uncertainties in the aircraft dynamics are assumed as follows:

$$F_d = 5\begin{bmatrix} 2\sin(3t) + \cos(t) \\ \sin(3t) + 2\cos(t) \\ 0.5\sin(3t) + 3\cos(t) \end{bmatrix} + \begin{bmatrix} \delta_{p1} \\ \delta_{p2} \\ \delta_{p3} \end{bmatrix}, \tau_d = 2\begin{bmatrix} 0.5\sin(3t) + 0.8\cos(t) \\ 0.5\sin(3t) + 0.5\cos(t) \\ 2\sin(3t) + 0.5\cos(t) \end{bmatrix} + \begin{bmatrix} \delta_{a1} \\ \delta_{a2} \\ \delta_{a3} \end{bmatrix}$$

where $\delta_{p1}$, $\delta_{p2}$, $\delta_{p3}$, $\delta_{a1}$, $\delta_{a2}$ and $\delta_{a3}$ are high-frequency noise.

The results from the presented controller simulation are seen in Figure 16 for hover-to-level and Figure 17 for level-to-hover. Although uncertainties are external disturbances exist in the dynamic equations of the tail-sitter aircraft, the controller approaches the desired trajectories and attitudes for both transition modes. We can carry out that the thrusts generated by rotors during forward flight mode (55N) are far smaller than that during hover (500N). Therefore, under the same cruising velocity, the presented tail-sitter aircraft can save much energy than helicopter. This can increase endurance cruising time and flying distance. Furthermore, the computational analysis and simulations exhibit the agile maneuverability of the presented tail-sitter aircraft with simple control algorithm.

| Parameter | Value | Parameter | Value | Parameter | Value | Parameter | Value |
|---|---|---|---|---|---|---|---|
| $m$ | $50kg$ | $g$ | $10m/s^2$ | $k_{1,1}$ | 5 | $k_{1,2}$ | 10 |
| $\rho$ | $1.225kg/m^3$ | $l_1$ | $0.6m$ | $k_{2,1}$ | 4 | $k_{2,2}$ | 6 |
| $l_2$ | $1m$ | $l_3$ | $0.8m$ | $k_{3,1}$ | 6 | $k_{3,2}$ | 8 |
| $l_a$ | $0.08m$ | $S$ | $0.45m^2$ | $k_{4,1}$ | 6 | $k_{4,2}$ | 11 |
| $J_{xb}$ | $0.2m^2kg$ | $J_{yb}$ | $0.2m^2kg$ | $k_{5,1}$ | 3 | $k_{5,2}$ | 7 |
| $J_{zb}$ | $0.4m^2kg$ | $S_i$ | $0.04m^2$ | $k_{6,1}$ | 6 | $k_{6,2}$ | 11 |
| $C_{L0}$ | 0.32 | $C_{D0}$ | 0.008 | $\lambda_c$ | 0.2673 | $K$ | 6 |
| $C_{L\alpha}$ | 0.5 | $C_{L\delta}$ | 0.05 | $k_{p1}$ | 0.2 | $k_{p2}$ | 0.6 |
| $J_r$ | $0.01m^2kg$ | $b$ | $5\times10^{-4}$ | $k_{a1}$ | 0.8 | $k_{a2}$ | 0.5 |
| $k$ | $3\times10^{-5}$ | $C_{f\delta}$ | 0.02 | $C_{lf\alpha}$ | 0.0802 | $C_{df0}$ | 0.0063 |
| $k_{uv}$ | 0.4376 | $k_f$ | 2.6583 | $C_{df\alpha}$ | 0.0094 | $\delta_a$ | $0.13686rad$ |
| $C_{\phi i}$ | 0.02 | $C_{\psi i}$ | 0.01 | $\alpha_{d0}$ | $5°$ | $\gamma_s$ | $80°$ |

Table 1 Parameters of aircraft model and flight

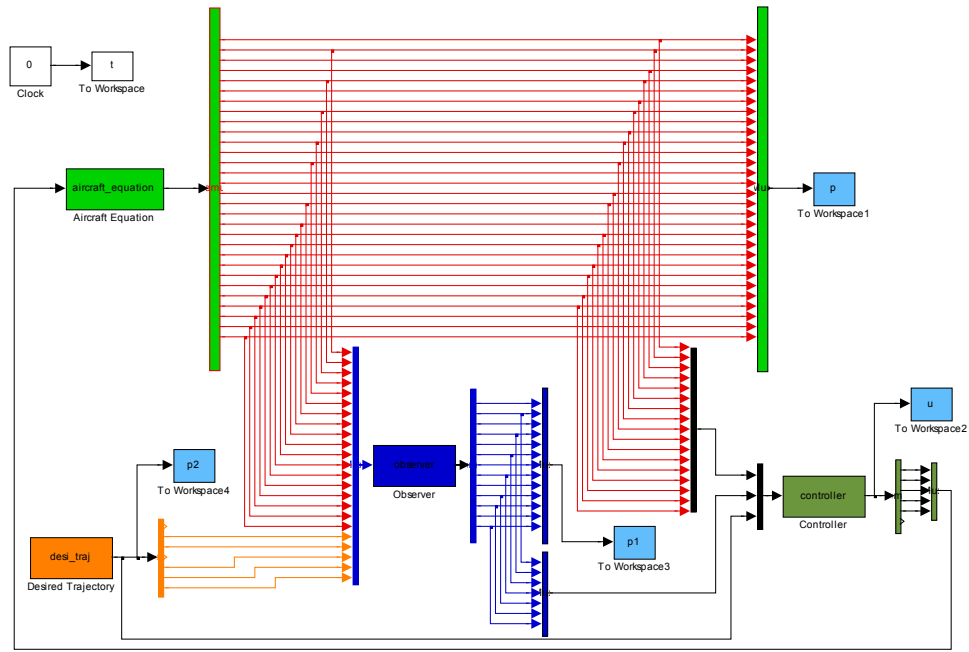

Figure 15 Simulink of aircraft control system

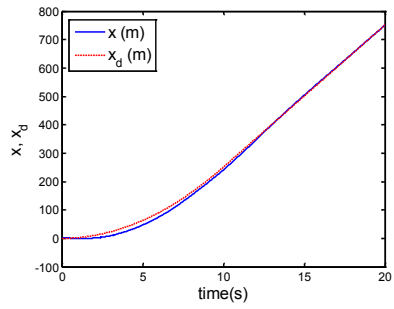
(a) Position in *x* direction

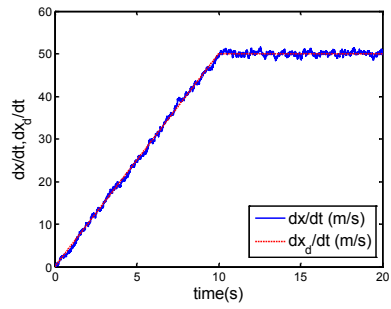
(b) Velocity in *x* direction

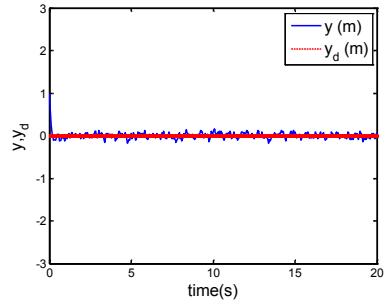
(c) Position in *y* direction

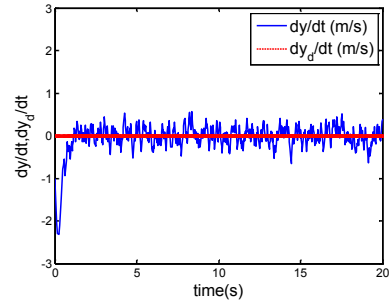
(d) Velocity in *y* direction

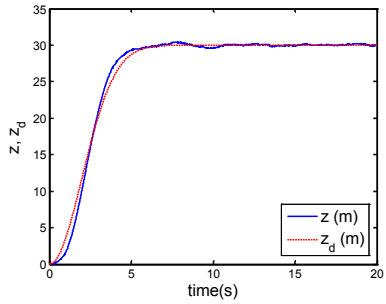
(e) Position in *z* direction

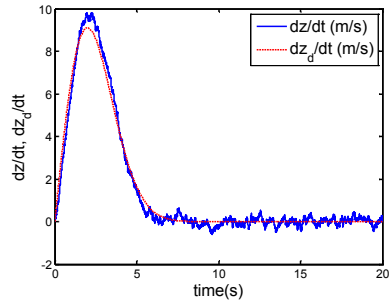
(f) Velocity in *z* direction

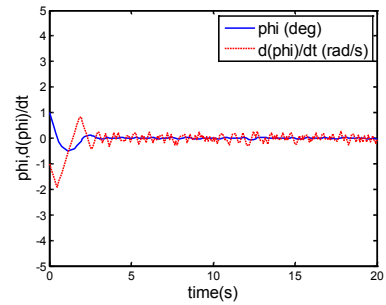
(g) roll angle and roll rate

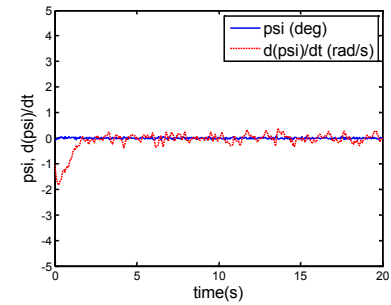
(h) yaw angle and yaw rate

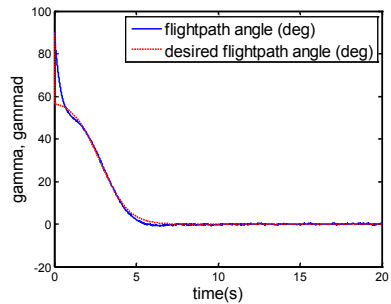
(i) Flightpath angle

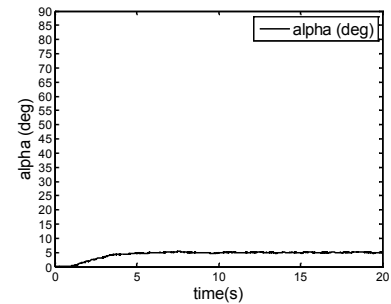
(j) Angle of attack

Figure 16 Mode transition from hover to forward flight

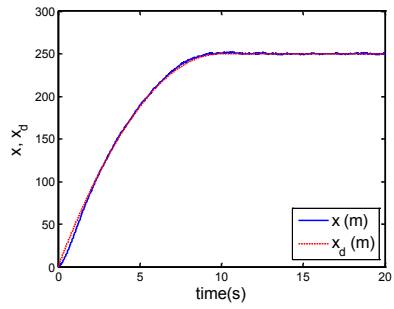
(a) Position in *x* direction

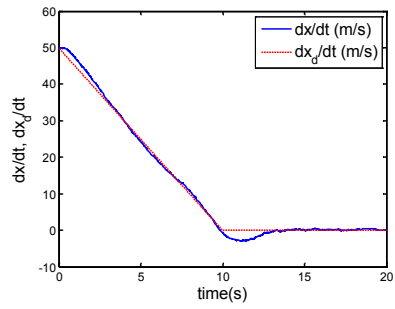
(b) Velocity in *x* direction

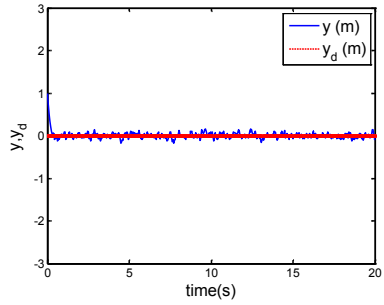
(c) Position in *y* direction

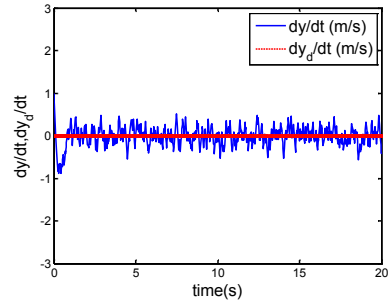
(d) Velocity in *y* direction

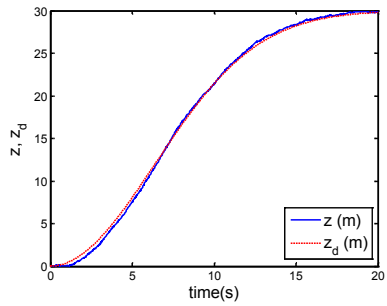
(e) Position in *z* direction

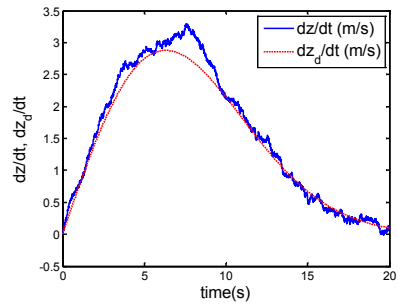
(f) Velocity in *z* direction

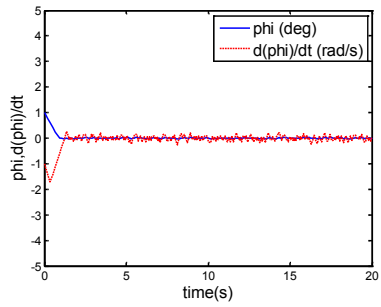
(g) roll angle and roll rate

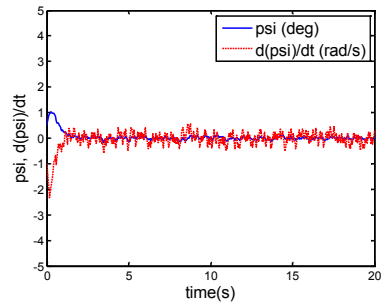
(h) yaw angle and yaw rate

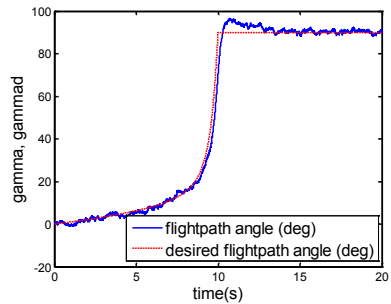
(i) Flightpath angle

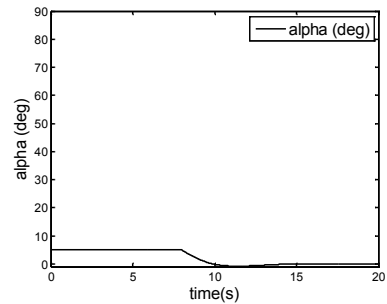
(j) Angle of attack

Figure 17 Mode transition from forward flight to hover

# 8 CONCLUSIONS

In this paper, a novel model of an agile tail-sitter aircraft is presented. Not only the aircraft can hover, take off and land vertically, but also the forward flight with high speed be implemented. Comparing with the conventional tail-tail aircraft, more agile maneuverability can be obtained. Moreover, the aircraft is controlled easily to implement the mode transitions. Our future work is to implement the hardware of the presented tail-sitter aircraft.

# 9 APPENDIX

**Proof of Theorem 1:**

Let the radius of co-axial blade be $R_r$. The effect of the root cut out (the inner, non-aerodynamic portion of the blade) can be estimated. If $l_a$ is the non-dimensional radius of the root cut-out, then the effective area becomes

$$A = \pi R_r^2 - \pi l_a^2 \tag{109}$$

The mass flow rate through the upper rotor is

$$\dot{m}_u = \rho A U_u \tag{110}$$

where $U_u = ((V_b \sin\alpha)^2 + (V_b \cos\alpha + v_u)^2)^{1/2}$. Therefore,

$$\dot{m}_u = \rho A ((V_b \sin\alpha)^2 + (V_b \cos\alpha + v_u)^2)^{1/2} \tag{111}$$

Thus, the thrust on the upper rotor may be written as

$$F_{cu} = \dot{m}_u (V_b \cos\alpha + w_u) - \dot{m}_u V_b \cos\alpha = \dot{m}_u w_u = 2\dot{m}_u v_u \tag{112}$$

and the power produced by upper rotor is

$$P_{cu} = T_u (V_b \cos\alpha + v_u) = 0.5\dot{m}_u (V_b \cos\alpha + w_u)^2 - 0.5\dot{m}_u V_b^2 \cos^2\alpha = 0.5\dot{m}_u (2V_b w_u \cos\alpha + w_u^2) \tag{113}$$

Therefore,

$$w_u (V_b \cos\alpha + v_u) = 0.5(2V_b w_u \cos\alpha + w_u^2) \tag{114}$$

or simply $w_u = 2v_u$.

The vena contracta of the upper rotor is an area of $A/2$ with velocity $V_b\cos\alpha+2v_u$. Therefore, at the plane of the lower rotor there is a velocity of $V_b\cos\alpha+2v_u+v_l$ over the inner one-half of the disk area (See Figure 11).

The mass of flow rates over the inner and outer parts of the lower rotor are $\dot{m}_{in} = 0.5\rho A \left( (V_b \cos\alpha + 2v_u + v_l)^2 + V_b^2 \sin^2\alpha \right)^{1/2}$ and $\dot{m}_{out} = 0.5\rho A \left( (V_b \cos\alpha + v_l)^2 + V_b^2 \sin^2\alpha \right)^{1/2}$, respectively. Therefore, the mass of flow rates over the lower rotor is

$$\dot{m}_l = \dot{m}_{in} + \dot{m}_{out} = 0.5\rho A \left( (V_b \cos\alpha + 2v_u + v_l)^2 + V_b^2 \sin^2\alpha \right)^{1/2} + 0.5\rho A \left( (V_b \cos\alpha + v_l)^2 + V_b^2 \sin^2\alpha \right)^{1/2} \tag{115}$$

The momentum flow out of plans 5 is $\dot{m}_l (V_b \cos\alpha + w_l)$. Thus, the thrust on the lower rotor may be determined as

$$F_{cl} = \dot{m}_l (V_b \cos\alpha + w_l) - \dot{m}_u 2v_u - \dot{m}_{out} V_b \cos\alpha \tag{116}$$

The work produced by the lower rotor is

$$P_{cl} = F_{cl} (V_b \cos\alpha + v_u + v_l) \tag{117}$$

and

$$F_{cl}(V_b \cos\alpha + v_u + v_l) = \frac{1}{2}\dot{m}_l(V_b\cos\alpha + w_l)^2 - \frac{1}{2}\dot{m}_{in}(V_b\cos\alpha + 2v_u)^2 - \frac{1}{2}\dot{m}_{out}V_b^2\cos^2\alpha \tag{118}$$

Assuming the co-axial is operated at equal power, i.e.,

$$P_{cu} = P_{cl} \tag{119}$$

Therefore, from (113), (117), (118) and (119), we obtain

$$\begin{aligned} F_{cl}(V_b\cos\alpha + v_u + v_l) &= 0.5\rho A(V_b\cos\alpha + v_u + v_l)(V_b\cos\alpha + w_l)^2 \\ &\quad - 0.5\rho\frac{A}{2}(V_b\cos\alpha + 2v_u)^3 - 0.5\rho\frac{A}{2}(V_b\cos\alpha + v_l)V_b^2 \\ &= 2\rho A(V_b\cos\alpha + v_u)^2 v_u \end{aligned} \tag{120}$$

For the Equations (116)-(120), it is difficult to solve for the direct relation between $v_u$ and $v_l$, i.e.,

$$v_l = \Phi(v_u,\alpha,V_b) \tag{121}$$

The fist-order Taylor expansion for (121) at $V_b = 0$ and $\alpha = 0$ is obtained as follow:

$$v_l = \Phi(v_u,\alpha,V_b) = k_{uv}v_u + \Gamma_{uv}(\alpha,V_b) \tag{122}$$

where $\Gamma_{uv}(\alpha,V_b)$ is the bounded function of $V_b$, and $\Gamma_{uv}(0,0) = 0$; $k_{uv}$ is a constant coefficient. When $V_b = 0$ and $\alpha = 0$, i.e., the co-axial propellers are in the hover. From (113), (116), (117) and (120), we obtain

$$w_l = 4(v_u + v_l)v_u/(2v_u + v_l) \tag{123}$$

and

$$2(v_l/v_u)^3 + 5(v_l/v_u)^2 + 2(v_l/v_u) - 2 = 0 \tag{124}$$

Solving Equation (124) provides

$$v_l = 0.4376 v_u \tag{125}$$

From (122), $k_{uv} = 0.4376$ is obtained.

It is known that there exists a positive non-dimensional quantity $\lambda_c$, which is called the induced inflow ratio, such that

$$v_u = \lambda_c R_d \omega_u,\ v_l = \lambda_c R_d \omega_l \tag{126}$$

Therefore, the following expression holds:

$$\omega_l = \frac{v_l}{v_u}\omega_u = \frac{k_{uv}v_u + \Gamma_{uv}(\alpha,V_b)}{v_u}\omega_u = k_{uv}\omega_u + \Gamma_{uv}(\alpha,V_b)\omega_u/v_u \tag{127}$$

Then, $\Gamma_\omega(\alpha,V_b) = \Gamma_{uv}(\alpha,V_b)\omega_u/v_u$. From (112) and (126), it follows that

$$F_{cu} = 2\rho A v_u^2 + 2\rho A v_u V_b = 2\rho A \lambda_c^2 R_d^2 \omega_u^2 + 2\rho A v_u V_b \tag{128}$$

And from (116) and (118), we obtain

$$w_l = 2.3590 v_u + \Gamma_{wv}(\alpha,V_b) \tag{129}$$

where $\Gamma_{wv}(\alpha,V_b)$ is the bounded function of $V_b$, and $\Gamma_{wv}(0,0) = 0$. Therefore, from (116), (127) and (129), it is obtained that

$$F_{cl} = 1.3913\rho A v_u^2 + \Gamma_{Fcl}(\alpha,V_b) = 1.3913\rho A \lambda_c^2 R_d^2 \omega_u^2 + \Gamma_{Fcl}(\alpha,V_b) \tag{130}$$

Finally, we obtain

$$F_c = F_{cu} + F_{cl} = \bar{k}_u \omega_u^2 + \Gamma_{Fc}(\alpha,V_b) \tag{131}$$

where $\bar{k}_u = 3.3913\rho A \lambda_c^2 R_d^2$, $\Gamma_{Fc}(\alpha,V_b) = 2\rho A v_u V_b + \Gamma_{Fcl}(\alpha,V_b)$.

This concludes the proof. ∎

**Proof of Theorem 2:**

For system (49) and observer (50), let

$$\hat{e}_{i1} = \hat{\zeta}_{i1} - \zeta_{i1}, \hat{e}_{i2} = \hat{\zeta}_{i2} - \zeta_{i2} \tag{132}$$

The system error between (50) and (49) is

$$\dot{\hat{e}}_{i1} = e_{i2} - k_{i,1}|\hat{e}_{i1}|^{1/2}\operatorname{sign}(\hat{e}_{i1}) \tag{133}$$
$$\dot{\hat{e}}_{i2} = -k_{i,2}\operatorname{sign}(\hat{e}_{i1}) - \eta_i(t)$$

Select the Lyapunov function be

$$V_i = \varsigma^T P_a \varsigma \tag{134}$$

where $\varsigma = \begin{bmatrix} |\hat{e}_{i1}|^{\frac{1}{2}}\operatorname{sgn}(\hat{e}_{i1}) & \hat{e}_{i2} \end{bmatrix}^T$, and $P_a$ is a positive definite and symmetrical matrix with the following form:

$$P_a = \frac{1}{2}\begin{bmatrix} 4k_{i2} + k_{i,1}^2 & -k_{i,1} \\ -k_{i,1} & 2 \end{bmatrix} \tag{135}$$

Differentiating $V_i$ with respect to time yields

$$\dot{V}_i \leq -\left[c_a/(\lambda_{\min}\{P\})^{1/2}\right]V_i^{1/2} \tag{136}$$

where $c_a$ is a positive constant. From the definition of finite-time stability [39, 40], the system error (133) is finite-time convergent. This concludes the proof. ∎

**Proof of Theorem 3:**

From (72), (73) and (74), and after taking the time derivative of (75), we obtain

$$\ddot{e} = \frac{1}{2}\dot{M}_q \tilde{\Omega}_\Lambda + \frac{1}{2}M_q \dot{\tilde{\Omega}}_\Lambda \tag{137}$$
$$= \frac{1}{2}\dot{M}_q \tilde{\Omega}_\Lambda + \frac{1}{2}M_q \left[-J^{-1}(\tilde{\Omega}_\Lambda + \Omega_{\Lambda d}) \times J(\tilde{\Omega}_\Lambda + \Omega_{\Lambda d}) + J^{-1}\tau_r + J^{-1}(\tau_w + \tau_{gyro}) + J^{-1}\tau_d - \dot{\Omega}_{\Lambda d}\right]$$

Considering controller (76), the closed-loop error system for the attitude dynamics is

$$\ddot{e} = -k_{a1}e - k_{a2}\dot{e} + \frac{1}{2}M_q\left[J^{-1}\tau_d - J^{-1}\hat{\tau}_d\right] \tag{138}$$

For $t \geq t_s$, selecting the Lyapunov function to be $V_a = k_{a1}e^T e_p + (1/2)\dot{e}^T \dot{e}_p$, we can obtain that $e \to 0$ and $\dot{e} \to 0$ as $t \to \infty$. This concludes the proof. ∎

**Proof of Theorem 4:**

In the light of Theorem 2, for $t \geq t_s$, the observation signals $\frac{1}{m}R\hat{\bar{F}}_d = \frac{1}{m}R\bar{F}_d$. Considering controller (79), the closed-loop error system for position error dynamics (77) is

$$\dot{e}_{\Gamma 1} = e_{\Gamma 2}$$
$$\dot{e}_{\Gamma 2} = -k_{p1}e_{\Gamma 1} - k_{p2}e_{\Gamma 2} + [\frac{1}{m}R\overline{F}_d - \frac{1}{m}R\widehat{\overline{F}}_d] \qquad (139)$$

For $t \geq t_s$, selecting the Lyapunov function to be $V_p = k_{p1}e_{\Gamma 1}^{\mathrm{T}}e_{\Gamma 1} + (1/2)e_{\Gamma 2}^{\mathrm{T}}e_{\Gamma 2}$, we can obtain that $e_{\Gamma 1} \to 0$ and $e_{\Gamma 2} \to 0$ as $t \to \infty$. This concludes the proof. ∎

## 10 LIST OF SYMBOLS

$F_c$ sum of the thrusts of the co-axial propellers

$F_{ri}, i = 1, \cdots, 4$ thrust of each quad rotor

$F_d$ forces due to uncertainties and external disturbances

$\tau_{ri}, i = 1, \cdots, 4$ reactive torque generated in free air by the rotor due to rotor drag for quad rotors

$\tau_{gyro}$ gyroscopic effects of the propellers

$\tau_d$ moments due to the uncertainties and external disturbances

$f_{ai}, i = 1, \cdots, 4$ force generated by each blade of torque amplifier

$\omega_i, i = 1, \cdots, 4$ rotational velocity of each quad rotor

$\omega_u$ rotational velocity of upper rotor

$\omega_l$ rotational velocity of lower rotor

$m$ mass of aircraft

$g$ acceleration of gravity

$l_1$ distance between center of gravity of aircraft and force operating point of fixed wing

$l_2$ distance between center of gravity of aircraft and plane center of quad rotors

$l_3$ distance between two quad rotor

$l_c$ distance between center of gravity of aircraft and fixed wing

$\phi$ roll angle

$\theta$ pitch angle

$\psi$ yaw angle

$\alpha$ angle of attack

$\gamma$ flightpath angle

$L_1$ lift force generated by left fixed wing

$L_2$ lift force generated by right fixed wing

$L_f$ the lift force generated by the fuselage, respectively

$D_1$ drag force generated by left fixed wing

$D_2$ drag force generated by right fixed wing

$D_f$ the drag force generated by the fuselage, respectively

$S$ area of fixed wing

$C_{L0}$ lift coefficient when angle of attack $\alpha$ is equal to zero for fixed wing

$C_{L\alpha}$ lift coefficient due to angle of attack $\alpha$

$C_{lf}$ lift coefficient of fuselage

$C_{df}$ the drag coefficient of fuselage

$C_{df0}$ the constant in the coefficients of drag force of fuselage

$\delta_i$ normal flap bias angle of fixed wing

$C_{L\delta}$ lift coefficient due to flap bias angle $\delta_i$

$J$ inertial matrix of aircraft

$p_\Gamma$ position of center of gravity relative to right handed inertial frame

$\upsilon_\Gamma$ velocity vector of center of gravity relative to right handed inertial frame

$F_f$ body force of aircraft

$R$ transformation matrix representing the orientation of the rotorcraft

$\tau$ sum of the moments in the fixed-body frame

$\Omega_A$ angular velocity of the airframe expressed in body frame

$\Omega_\Gamma$ angular velocity of the airframe expressed in inertial frame

$V_b$ velocity of center of gravity relative to right handed inertial frame

$v_u$ induced velocity through upper disk

$v_l$ induced velocity through lower disk

$w_u$ velocity of vena contracta of the upper rotor

$w_l$ velocity of vena contracta of the lower rotor

$v_i, i = 1,\cdots,4$ induced velocity of each quad rotor

$w_i, i = 1,\cdots,4$ velocity of vena contracta of each quad rotor

$P_{ri}, i = 1,\cdots,4$ power for each quad rotor

$F_{cu}$ thrust on the upper rotor

$F_{cl}$ thrust on the lower rotor

$l_a$ radius of the root cut-out of the co-axial

$A$ effective area of co-axial disk

$A_q$ area of each quad rotor disk

$R_d$ the radius of quad rotor disk

$R_r$ the radius of co-axial blade

$\rho$ air density

$\dot{m}_u$ mass flow rate through the upper rotor

$\dot{m}_l$ mass of flow rates over the lower rotor

$\dot{m}_{in}$ mass of flow rates over the inner part of the lower rotor

$\dot{m}_{out}$ mass of flow rates over the outer part of the lower rotor

$P_{cu}$ power produced by upper rotor

$P_{cl}$ power produced by lower rotor

$v_e$ the induced velocity for each rotor operating in isolation

$P_{ce}$ power for each rotor in isolation

$\kappa_{int}$ induced power factor of co-axial

$\kappa_{inth}$ induced power factor for all the propellers

$\lambda_c$ induced inflow ratio

$b$ force coefficient of one of quad rotors

$k$ torque coefficient of one of quad rotors

$J_r$ the moment of inertia of each rotor

$\tau_{ai}, i=1,\cdots,4$ torque generated by each blade of torque amplifier

$S_f$ blade area of torque amplifier

$\delta_{fi}$ normal flap bias angle of torque amplifier

$C_{f\delta}$ lift coefficient created by the flap bias angle $\delta_{fi}$.

$C_{Df}$ drag coefficient of blade area of torque amplifier

$D_{fi}, i=1,\cdots,4$ drag force generated by each blade of torque amplifier

$k_f$ force coefficient of torque amplifier

$G_b$ gyroscopic torques due to the combination of the rotation of the airframe and quad rotors